\documentclass[conference]{IEEEtran}

\IEEEoverridecommandlockouts

\usepackage{setspace}
\usepackage{enumerate}
\usepackage{cite}
\usepackage{times}
\usepackage{url}
\usepackage{graphicx}
\usepackage{subfigure}
\usepackage{amsmath,amssymb}

\usepackage{amssymb,amsmath,amsfonts,graphicx,epsfig}
\usepackage{color,psfrag,colordvi,graphics}
\usepackage{mathptmx}  
\usepackage{dcolumn}

\newcommand{\dive}{\nabla\cdot}

\newcommand{\bu}{{\bf u}}
\newcommand{\bv}{{\bf v}}
\newcommand{\bw}{{\bf w}}

\newcommand{\bff}{{\bf f}}

\def\bnu{{\bf \nu}}
\def\btau{{\bf \tau}}

\def\ds{\displaystyle}

\newcommand{\be}{{\bf e}}
\newcommand{\we}{{\text{We}}}
\newcommand{\re}{{\text{Re}}}
\newcommand{\fr}{{\text{Fr}}}
\begin{document}
\title{On the Navier-slip boundary condition for computations of impinging droplets}

\author{
\IEEEauthorblockN{
Jagannath Venkatesan\IEEEauthorrefmark{1},
Sashikumaar Ganesan\IEEEauthorrefmark{2}
}
\IEEEauthorblockA{Numerical Mathematics and Scientific Computing\\
Supercomputer~Education~and~Research~Centre\\
Indian Institute of Science, Bangalore, India\\
\IEEEauthorrefmark{1}jagan@nmsc.serc.iisc.in,
\IEEEauthorrefmark{2}sashi@serc.iisc.in\\}
}

\maketitle

\begin{abstract}
 A mesh-dependent relation for the slip number in the Navier-slip with friction boundary condition for computations of impinging droplets with sharp interface methods is proposed. 
 The relation is obtained as a function of Reynolds number, Weber number and the mesh size. 
 The proposed relation is validated for several test cases by comparing the numerically obtained wetting diameter with the experimental results. 
 Further, the computationally obtained maximum wetting diameter using the proposed slip relation is  verified with the theoretical predictions. 
 The relative error between the computationally obtained maximum wetting diameter and the theoretical predictions is less than 10\% for impinging droplet on a hydrophilic surface, and the error increases in the case of hydrophobic surface.  
\end{abstract}


\section{Introduction}
\label{intro}

Impinging droplets are encountered in many scientific and industrial applications such as spray cooling, inkjet printing, fuel injecting, etc. 
 Simulating such flows is complicated by the violation of the no-slip condition in the vicinity of the moving contact line, where the liquid-solid, solid-gas interfaces and the free surface intersect. 
 The choice of the classical hydrodynamic ``no-slip'' boundary condition in the neighbourhood of the moving contact line leads to an unsatisfactory model that induce multivalued velocity field, refer~\cite{DUS76, HOC76, HUH71, WJ80}.
 To alleviate this problem, often the contact line is  allowed to move  instead of imposing  zero fluid velocity at the contact line. 
 A number of approaches have been proposed in the literature to move the contact line. 
 In one of the approaches,  the velocity of the moving contact line is prescribed as a function of the local dynamic contact angle~\cite{DEG85}, which is the angle between the liquid-solid interface and the free surface. 
 Several models  for the  contact line velocity have been proposed in the literature, see Eggers et al.~\cite{EGG04} for an overview. 
 These models are mostly valid for wetting or perfectly wetting liquids. Further, the local dynamic contact angle is seldom available, and it varies for different flow configurations. 
 Therefore,  this approach is hardly used in computations.  
 Another approach is to allow the fluid in the vicinity of the contact line to slip over the solid surface, refer~\cite{DUS76, HOC77, HUH77} i.e., the relative velocity of the solid and liquid will be nonzero. 
 To induce a slip, the slip with friction boundary condition 
\begin{align} \label{slipbc}
 (\bw-\bu)\cdot\btau_S  = \epsilon_\mu\btau_S\cdot\mathbb{T}(\bu,{p})\cdot{\bnu_{S}}
\end{align}
is used, see  for example, Gennes~\cite{DEG85} and Ganesan~\cite{SG06}. The slip boundary condition has first been proposed by Navier~\cite{NAV1823}, and later studied by Kundt et al.~\cite{KUN1875} and Maxwell~\cite{MAX1879} for gas dynamics. 
Here, $(\bw-\bu)\cdot\btau_S$ is relative velocity (tangential) of the solid and the liquid,  and  $\btau_S\cdot\mathbb{T}(\bu,{p})\cdot{\bnu_{S}}$ is the shear stress of the liquid on the solid surface. 
Further,  $\epsilon_\mu$ is the slip coefficient which defines the extent to which the no-slip boundary is relaxed. 

A relation between the Greenspan slip coefficient and the grid-spacing of the numerical scheme has been proposed by Moriarty et al.~\cite{MORIA} for the moving contact line problem arising in dry wall coating.
A number of  theoretical and numerical investigations  have been performed by several authors for the choice of the slip coefficient for specific moving contact line problems.
Different expressions for the slip coefficient such as constants, functions of grid size, etc. have been proposed for specific moving contact line problems, refer~\cite{DUS76, HOC77,EGG04, CHO14, COX86, MAR12, REN07, THO97, MORIA, REN01, WE08, ZAL09}.
Molecular dynamics simulations were often used to predict the slip coefficient for moving contact line problems, see~\cite{HUA08,REN07}.
In almost all of these simulations, the moving contact line is considered in  channel flows. 
Hence, the predicted slip values  may not be generalized to all moving contact line problems, in particular, to impinging droplets. 
Even though the Navier-slip boundary condition~\eqref{slipbc} has been widely accepted for computations of moving contact line problems,  a general  mathematical expression or  an empirical correlation for the slip coefficient is not available for impinging droplet simulations.
The   slip coefficient value need not be same for a droplet impinging on a same surface with different impact velocities. 
Often the slip coefficients for impinging droplets were identified on an ad hoc basis by comparing the numerical results with the experiments, see~\cite{SG06,GAN13M,GRT14,GVR15}.
The wetting diameter of the droplet has been used as a key parameter to identify the appropriate slip coefficient. 
A smaller value of the slip coefficient will reduce the wetting  diameter, whereas a larger value increases the wetting diameter. 
Even though a deviation in the wetting diameter from the original value will induce a completely different flow dynamics in the droplet, the equilibrium state of the droplet is not affected by the slip coefficient. 
However, an appropriate choice of the slip coefficient has to be used in computations in order to obtain physically accepted numerical predictions, especially, the dynamics of the fluid flow during the droplet deformation. 

It is the purpose of this paper to study the effect of the slip coefficient for different impact velocities and droplet sizes, and to compare the numerically obtained wetting diameter with experiments. 
Further, an  expression for the slip coefficient is proposed.  
Apart from the choice of the slip coefficient, the inclusion of the contact angle into the model is very challenging. 
In particular, the choice of the contact angle value is very important in computations of impinging droplets, see Ganesan~\cite{GAN13M} for a recent comparative study of different contact angle models.
It has been observed that the equilibrium contact angle model is preferred for sharp interface methods. 
In discretization based numerical schemes (finite difference or finite volume or finite element methods), the contact angle is incorporated as a surface force, refer~\cite{GAN13M}.
Therefore, the measured dynamic contact angle need not be equal to the prescribed contact angle in the surface force until the droplet attains its equilibrium state. 
Consequently, the imbalance in the surface force induces a non-zero tangential velocity, and it necessitates slippage of liquid in the vicinity of the contact line. 
The above argument is another justification for the application of slip boundary condition in computations of moving contact line problems.  

An accurate approximation of the curvature and an appropriate discretization of pressure are essential to suppress spurious velocities in computations of free surface and interface flows, refer~\cite{GMT}.
In  Eulerian approaches such as level set  and volume-of fluid methods, the free surface is not resolved by the computational mesh. 
Thus, an accurate calculation of the curvature and the conservation of mass  are very challenging. 
Even  though a separate surface mesh is used to explicitly represent the free surface in  the  front tracking method, the Navier--Stokes solver mesh does not resolve the free surface, and  therefore the inclusion of the surface force is  still challenging. 
Alternatively, the free surface is resolved using the arbitrary Lagrangian-Eulerian (ALE) approach. Since the free surface is explicitly tracked in ALE approach, the surface force can accurately be incorporated in computations. 
Further, the inclusion of the contact angle is straight forward, refer~\cite{GAN13M}.
Even though handling the topological changes is very difficult in the ALE approach, it is possibly the most accurate approach for computations of free surface and two-phase flows when there is no topological change. 
Since the focus of this paper is to identify an appropriate  expression for the  slip coefficient, droplet impingement without any splashing and/or breakage is considered. Hence, the ALE approach is preferred in this study. 

The paper is organized as follows. 
The mathematical model and its dimensionless form of the governing equations are presented in Section~2. 
The used finite element scheme is briefly discussed in Section~3. 
The convergence study and an array of computations for impinging droplets are presented in Section~4. 
Further, a relation for the slip coefficient is derived and validated in this section. 
Finally, the findings are summarized in Section~5.

\section{Mathematical model}
\label{s2}
\begin{figure}[h]
\begin{center}
\unitlength5mm
\begin{picture}(6,11)
\put(3,6.0){\makebox(0,0){\includegraphics[height=6cm]{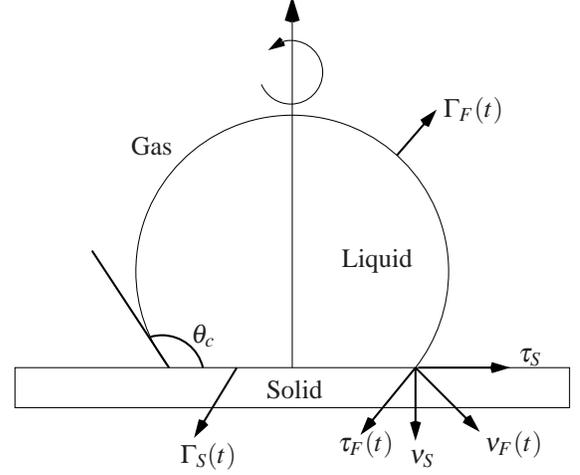}}}
\put(0 ,-0.5){$\Gamma_S(t)$}
\put(7 ,8.7){$\Gamma_F(t)$}
\put(4.2 ,-0.2){$\btau_F(t)$}
\put(9, 2.2){$\btau_S$}
\put(6.1 ,-0.5){$\bnu_S$}
\put(8.1 ,-0.2){$\bnu_F(t)$}
\put(0.3 ,2.7){$\theta_c$}
\put(2.3 ,1.2){Solid}
\put(4.3 ,4.7){Liquid}
\put(-1.3 ,7.7){Gas}
\end{picture}
\end{center}
\caption{Computational model of a droplet impinging on a horizontal surface.\label{Imp_Drop}}
\end{figure}

 We consider a spherical liquid droplet impinging on a horizontal surface, and the computation starts when the droplet comes into contact with the solid surface. 
 Computations are performed until the prescribed time or until the droplet comes into the equilibrium after spreading and recoiling.
 A schematic representation of the computational model is presented in Figure~\ref{Imp_Drop}. 
The liquid-solid interface and the free surface are represented by $\Gamma_S$ and $\Gamma_F$, respectively. 
Here, $\theta_c$ denotes the contact angle, $\btau_F$, $\bnu_{F}$ are unit tangential and unit outward normal vectors on $\Gamma_F$ and $\btau_S$, $\bnu_{S}$ are unit tangential and unit outward normal vectors on $\Gamma_S$, respectively.

\subsection{Governing Equations} 
The sequence of spreading and recoiling of an impinging liquid droplet is described by the time-dependent incompressible Navier-Stokes equations in a time-dependent domain $\Omega(t) \subset$ $\mathbb{R}^3$, $t$ $\in$ $(0,I)$.
\begin{equation}\label{NSE}
\begin{array}{rcll}
\displaystyle
  \frac{\partial \bu}{\partial t} + (\bu\cdot\nabla)\bu - \ds\frac{1}{\rho}\dive \mathbb{T}(\bu,{p}) &=& \bff  \quad \text{in } \Omega(t) \times (0,I) \\  
  \dive \bu &=& 0 \quad \text{in } \Omega(t) \times (0,I)
\end{array}
\end{equation}
where $\bu$ denotes the velocity of the fluid, p the pressure, $\rho$ the density, $I$ the given end time and $\bff$~=~(0,0,$-g$) the body force with gravitational constant $g$. 
The stress tensor $\mathbb{T}$ and the deformation tensor $\mathbb{D}$ for an incompressible Newtonian fluid are given by 
\begin{align*}
  \displaystyle
\mathbb{T}(\bu,{p}) := 2\mu \mathbb{D}(\bu)- p\mathbb{I}, \quad \mathbb{D}(\bu) = \frac{1}{2} \left(\nabla \bu + \nabla \bu^{T} \right),
\end{align*} 
where $\mu$ is the dynamic viscosity and $\mathbb{I}$ is the identity tensor.

\subsection{Initial and Boundary Conditions}
At time t~=~0, we assume that the droplet is of spherical shape with diameter d$_0$, and the initial velocity $\bu(x,0)$~=~(0,0,-$u_{imp}(x)$), where $u_{imp}$ is the impinging speed of the droplet.  
As mentioned in the introduction the Navier--slip with friction boundary condition is imposed on the liquid--solid interface and it reads
\begin{eqnarray*}
 \ds\bu\cdot\bnu_{S}~=& 0 \quad&\text{on } \Gamma_S(t) \times (0,I) \\
\btau_S\cdot\mathbb{T}(\bu,{p})\cdot{\bnu_{S}}~=& -\ds{\frac{1}{\epsilon_\mu}} \bu\cdot\btau_S \quad &\text{on }\Gamma_S(t) \times (0,I)
\end{eqnarray*}
The first condition is the no penetration boundary condition, i.e., the fluid cannot penetrate an impermeable solid and thus the normal component of the velocity is zero.  
The second condition is the slip with friction boundary condition, i.e., on the liquid-solid interface, the tangential stress is proportional to the tangential velocity of the fluid.  
Along the free surface, the force balancing condition
\begin{align*}
 \mathbb{T}(\bu,{p})\cdot{\bnu_{F}} = \nabla_{\Gamma_F}\cdot\mathbb{S}_{\Gamma_F} \quad \text{ on } \Gamma_F(t) \times (0,I)
\end{align*} 
is applied.  Here, $\nabla_{\Gamma_F}$ and $\nabla_{\Gamma_F}\cdot(\cdot)$ denote the tangential gradient  and tangential divergence, respectively, and are defined by 
\[
{\nabla}_{\Gamma_F}  (\cdot) = \mathbb{P}_{\bnu_F}\nabla (\cdot) , \qquad \nabla_{\Gamma_F}\cdot (\cdot)  = \text{tr }\left( \mathbb{P}_{\bnu_F}\nabla (\cdot) \right),
\]
where $\mathbb{P}_{\bnu_F}$~=~$\mathbb{I}-\bnu_F\otimes\bnu_F$ is the tangential projection.
The surface stress tensor, $\mathbb{S}_{\Gamma_F}$~\cite{SGJCP15} can be obtained as
\begin{align*}
\mathbb{S}_{\Gamma_F} = \sigma\,\mathbb{P}_{\bnu_F}.
\end{align*} 
Here, $\sigma$ is the surface tension. Further, the kinematic boundary condition
\begin{align*}
\bu \cdot \bnu_{F} =\bw \cdot \bnu_{F} \quad&\text{on }\Gamma_F(t) \times (0,I)
\end{align*} 
holds, i.e. the normal component of the fluid velocity on the free surface is equal to the normal component of the free surface velocity.

\subsection{Dimensionless form}
To write the Navier-Stokes equations in a dimensionless form, we introduce the scaling factors L and U as characteristic length and velocity, respectively. We define the dimensionless variables as 
\begin{align*}
 \tilde{x} = \frac{x}{L},  ~\tilde{u} = \frac{u}{U}, ~\tilde{t} = \frac{tU}{L},  ~\tilde{I} = \frac{IU}{L},  ~\tilde{p} = \frac{p}{\rho U^2}.
\end{align*} 
Using these dimensionless variables in the Navier-Stokes equations~(\ref{NSE}) and boundary conditions and omitting the tilde after-wards, we obtain the equations in a dimensionless form
$$
\begin{array}{rcll}
  \displaystyle
  \frac{\partial \bu}{\partial t} + (\bu\cdot\nabla)\bu - \ds\dive \mathbb{T}(\bu,{p}) &= &\ds\frac{1}{\text{Fr}}\be&  \text{in }\Omega(t)\\
  \dive \bu &= &0&  \text{in } \Omega(t) \\
\ds\bu\cdot\bnu_{S}&=& 0  &\text{on } \Gamma_S(t)  \\
\btau_S\cdot\mathbb{T}(\bu,{p})\cdot{\bnu_{S}}&=& -\ds{\beta_\epsilon} \bu\cdot\btau_S &\text{on }\Gamma_S(t) \\
\mathbb{T}(\bu,{p})\cdot{\bnu_{F}}&=&\ds\frac{1}{\text{We}}\nabla_{\Gamma_F}\cdot\mathbb{P}_{\bnu_F}& \text{on }\Gamma_F(t) \\
\quad \bu \cdot \bnu_{F}&=& \bw \cdot \bnu_{F} &\text{on }\Gamma_F(t) 
\end{array}
$$
where the dimensionless stress tensor is given by 
\begin{align*} 
 \mathbb{T}(\bu,{p}) = \ds\frac{2}{\text{Re}} \mathbb{D}(\bu)- p\mathbb{I}
\end{align*} 
and the Reynolds number, Froude number, Weber number and slip number are defined as
 \begin{align*} 
\text{Re} = \frac{\ds\rho U L}{\mu}, \quad
\text{Fr} = \ds\frac{U^2}{L g}, \quad
\text{We} = \ds\frac{\rho U^{2} L}{\sigma}, \quad
\beta_\epsilon = \ds\frac{1}{\epsilon_\mu \rho U}.
\end{align*}   
\section{Numerical scheme}
\label{s3}

We use finite element method together with the ALE approach to solve the governing equations. 
We first derive a weak form of the Navier--Stokes equations. 
And then, we briefly describe the ALE formulation.
After that, we discretize the weak problem in time and then in space. 
We briefly present the numerical scheme  here, and we refer to Ganesan et al.~\cite{SG06,GAN13M,GRT14,GVR15} for a detailed description.

\subsection{Weak formulation}
Let $L^2 (\Omega(t)$) and $H^{1} ( \Omega(t) )^3$ be the usual Lebesgue and Sobolev spaces. We define the velocity space V and pressure space Q as follows :
 \begin{align*} 
V &= \{ v \in H^{1} ( \Omega(t) )^3 :  \bv\cdot\bnu_{S} = 0  \text{ on }  \Gamma_S(t) \} \\
Q &=  \{ q \in L^{2} ( \Omega(t) ) \}
\end{align*} 
To derive a weak form of the time-dependent incompressible Navier-Stokes equations, we multiply the momentum and mass balance equations by the test functions v $\in$ V and q $\in$ Q, respectively and integrate over $\Omega(t)$. 
After applying the Gaussian theorem for the stress tensor term and incorporating the boundary conditions, the weak form of the Navier-Stokes equations read: \\

\noindent For given $\Omega$(0), $\bu$(x,0), find ($\bu$(x,t), p(x,t)) $\in$ $V~\times~Q$ such that 
 \begin{eqnarray}\label{weak}
\displaystyle
\left(\frac{\partial \bu}{\partial t},\mathbf v\right) + a(\hat{\bu},\bu,\bv) - b(p,\bv) + b(q,\bu) = f(\bv)
\end{eqnarray}
for all v $\in$ V and q $\in$ Q. Here, 
\begin{eqnarray*}
 \displaystyle a(\hat{\bu},\bu,\bv)&=&\ds\frac{2}{Re}\int_{\Omega(t)}\mathbb{D}(\bu):\mathbb{D}(\bv)\,dx + \ds\int_{\Omega(t)} (\hat{\bu}\cdot\nabla)\bu \bv \,dx  \quad \\
 &+& \ds{\beta_\epsilon}\int_{\Gamma_S(t)}(\bu\cdot\tau_{S})(\bv\cdot\tau_{S})\,d\gamma_S \\
b(q,\bv) &=& \int_{\Omega(t)} {q} \dive \bv \, dx\\
f(\bv)&=&\frac{1}{Fr}\int_{\Omega(t)}\be\cdot\bv\,dx-\frac{1}{We}\int_{\Gamma_F(t)}\mathbb{P}_{\bnu_F} : \nabla_{\Gamma_F}\bv d\gamma \quad \\
&+& \frac{1}{We} \int_{\gamma_{cl}}  \cos(\theta_c) \bv\cdot\tau_{S}~ds,
\end{eqnarray*}
where $\gamma_{cl}$ denotes the contact line.
We refer to Ganesan et al.~\cite{GRT14} for the inclusion of the contact angle.
The contact angle model: $\theta_c$~=~$\theta_e$ is used in all computations.
The choice of equilibrium value in computations  does not mean that the dynamic contact angle is fixed to the equilibrium value during the computations. 
Since the contact angle is included in the weak form as a natural boundary condition without imposing any condition on the geometry or on the contact-line velocity, the movement of the free surface in computations induces the hysteresis behaviour in the contact angle. 
A detailed investigation on the effects of different contact angle models has been studied in Ganesan~\cite{GAN13M}, and the equilibrium value is preferred for sharp interface methods.

\subsection{Arbitrary Lagrangian--Eulerian Approach}
Let $A_t$ be a family of mappings, which at each $t \in [0,I)$ maps a point (ALE coordinate) Y of a reference domain $\hat{\Omega}(t)$ onto the point (Eulerian coordinate) X of the current domain $\Omega(t)$:
 \begin{align*} 
A_{t}\colon{\hat{\Omega}(t)} \rightarrow\Omega(t), \qquad A_{t}(Y) = X(Y,t)
\end{align*} 
We assume that the mapping $A_t$ is homeomorphic, i.e., $A_{t}$ is invertible with continuous inverse. 
We also assume that the mapping is differentiable almost everywhere in $[0,I)$. 
The reference domain  $\hat\Omega(t)$ can simply be the initial domain $\Omega_0$ or the previous time-step domain when the deformation of the domain is large. 
Next, for a vector function $\bu\in C^0({{\Omega(t)}})$ on the Eulerian frame, we define their corresponding function $\hat \bu\in C^0({{\hat\Omega(t)} })$ on the ALE frame as
\begin{align*}
\hat \bu :\hat\Omega(t) \rightarrow \mathbb{R}, \quad
 \hat{\bu}:=\bu\circ A_t,\qquad \text{with} \quad \hat{\bu}(Y, t) = \bu(A_t(Y), t).
\end{align*}
Further, the time derivative of $\bu$  on the ALE frame is defined as
\begin{align*}
\ds\frac{\partial \bu}{\partial t} \Big|_{Y}:\Omega(t) \rightarrow \mathbb{R}, \quad
\ds\frac{\partial \bu  }{\partial t} \Big|_{Y }(X,t) = \ds\frac{\partial \hat \bu }{\partial t}(Y,t), \quad
 Y= A_t^{-1}(X).
\end{align*}
We now apply the chain rule to the time derivative of $\bu\circ A_t$ on the ALE frame to get
\begin{align*}
\ds\frac{\partial \bu}{\partial t} \Big|_{Y} = \ds\frac{\partial \bu}{\partial t} (X,t)
 + \ds\frac{\partial X}{\partial t}\Big|_{Y}\cdot\nabla_x \bu  = \ds\frac{\partial \bu}{\partial t}\Big|_{X}
 + \bw\cdot\nabla_x \bu,
\end{align*}
where $\bw$ is the domain velocity.
Using the above relation, we write the Navier-Stokes equations in the ALE form as
 \begin{align*} 
\frac{\partial \bu}{\partial t}\big|_Y - \dive\mathbb{T}(\bu, p) + ((\bu-\bw)\cdot\nabla)\bu = \bf{f}, \qquad \dive \bu  =  0.
\end{align*} 
Since the free surface is resolved by the computational mesh in the ALE approach, the spurious velocities if any can be suppressed when the surface force is incorporated into the scheme accurately as discussed in Ganesan et al.~\cite{GMT}. 
The application of ALE approach adds additional mesh velocity convective term in the model equations, and the mesh velocity needs to be computed at every time step. 

\subsection{Axisymmetric formulation}
The computational domain of the considered problem is time-dependent and a very fine discretization, both in space and in time is needed to get an accurate solution. 
This requirement increases the computational costs in 3D. 
Since the considered domain is rotational symmetric, a 2D geometry with 3D-axisymmetric configuration is used.  
Thus, we rewrite the volume and surface integrals in~(\ref{weak}) into area and line integrals as described in Ganesan et al.~\cite{GAN07}.
It allows to use two-dimensional finite elements for velocity and pressure.
Further, it reduces the computational complexity in mesh movement.
 
\subsection{Discretization in time and space}
Various time stepping methods have been proposed in the literature. 
The Euler schemes are of first order and the Crank-Nicolson is of second order but the latter is not strongly A-stable. 
Thus, we prefer the second order, strongly A-stable fractional-step scheme, refer~\cite{BGR, STK}.  
Next to guarantee the stability and high accuracy we prefer the inf-sup stable finite elements of second order.
We use triangular elements that approximates the complex domains more accurately. 
One of the popular inf-sup stable finite elements used in computations is the Taylor--Hood element, i.e., continuous piecewise quadratic approximations for the velocity and continuous piecewise linear for pressure, and it is used in this paper. 
Further, a fixed point iteration is used to linearize the Navier-Stokes equations at every time step. 
Finally, the system of linear algebraic equations arising from the linearized Navier-Stokes equations is solved using UMFPACK (direct solver), refer~\cite{Dav04a}. 

\subsection{Mesh movement}
A linear elastic mesh update technique is used to handle the mesh movement.
After solving the Navier-Stokes equations in each time step, the boundary displacement is calculated using the fluid velocity on the boundary.
Using the boundary displacement as a Dirichlet boundary condition, the linear elastic equation is then solved for the inner points displacement.
Finally, the mesh is moved with the computed displacement to get the next time step domain, see Ganesan et al.~\cite{GRT14} for more details.

\section{Numerical Results}\label{s4}
In this section, we present the numerical results for an axisymmetric spherical liquid droplet impinging on a horizontal surface. 
We first perform a mesh convergence study in which we vary the number of points on the free surface. 
After that, we perform an array of simulations for glycerin and water droplets impinging on a glass surface with different impinging velocities.
The flow dynamics of the droplet depends on the surface characteristics, Reynolds, Weber, Froude and the slip number. 
Among these numbers only the slip number is a numerical model parameter.
Thus, the effect of the slip number on the flow dynamics of droplet for different impinging velocities and liquids are studied. 
The appropriate slip number for each test case is identified by comparing the numerically obtained dimensionless wetting diameter with their corresponding experimental result presented in the literature. 
Based on the identified slip values, a correlation for the slip number in terms of the mesh size, the Reynolds and the Weber number is obtained. 
An array of simulations are performed by varying the equilibrium contact angle to check the applicability of the proposed slip relation for hydrophilic and hydrophobic surfaces. 
The maximum wetting diameter obtained from the simulations using the proposed slip relation are compared with the analytical values and other experiments to validate the proposed slip relation.    

\subsection{Mesh convergence study}
In this section we perform a mesh convergence study for the proposed numerical scheme.
Space discretization is a very important aspect in CFD simulations in order to obtain accurate numerical results.
Numerical simulation with extremely small mesh size is ideal to the continuum problem but it is not possible in practice due to the limited computational resources. 
We use open source package Triangle for mesh generation, which is based on constrained delaunay triangulation and the constraint we impose in our problem is the number of points used to track the free surface.
In order to identify a feasible mesh size, we perform an array of simulations with a test example by varying the number of points on the free surface.

\begin{figure*}
\begin{center}
\unitlength1cm
\begin{picture}(12,3.5)
\put(-1.2,-2.35){\makebox(6,8){\includegraphics[width=0.5\textwidth]{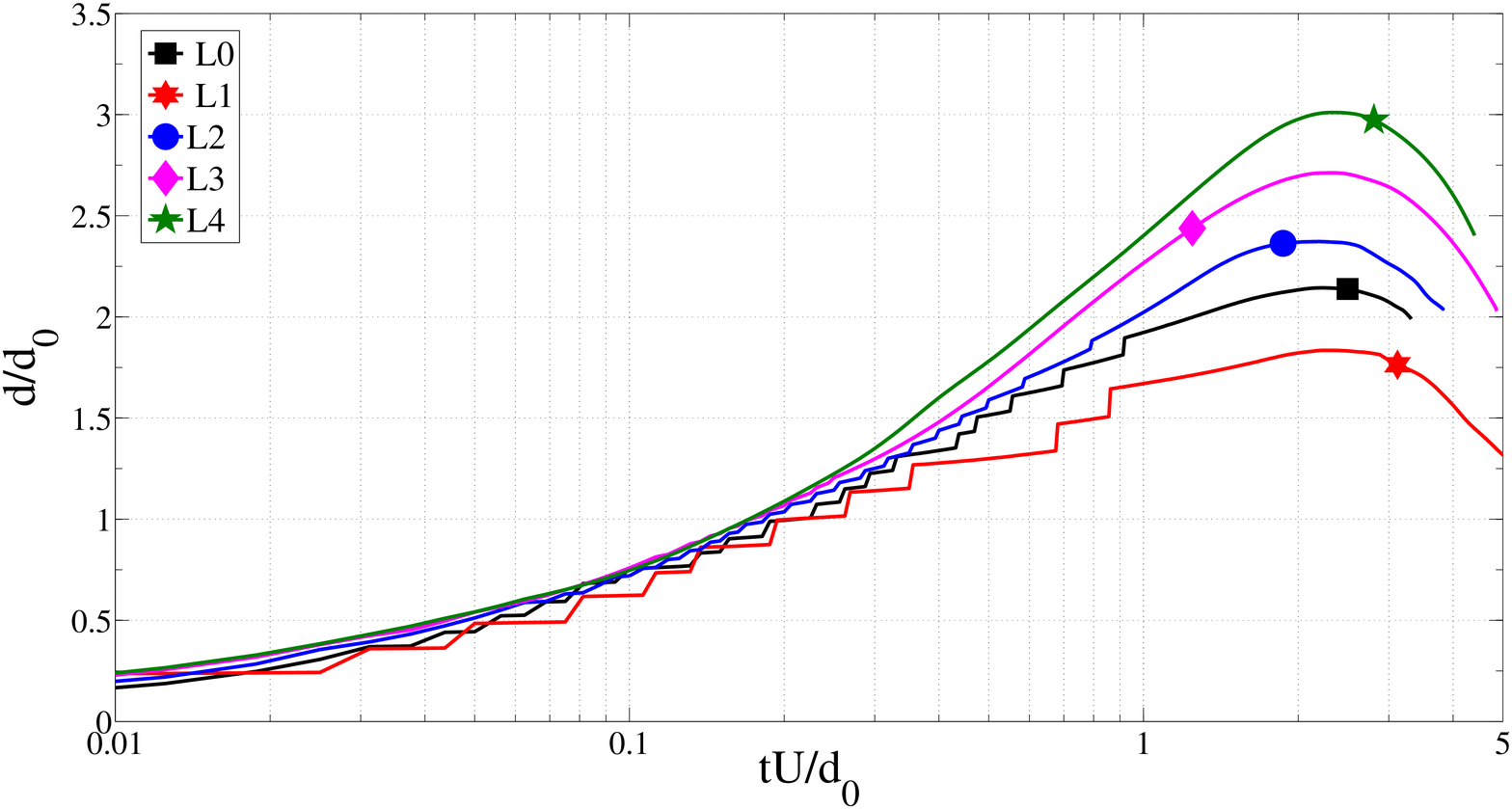}}}
\put(7.3,-2.35){\makebox(6,8){\includegraphics[width=0.5\textwidth]{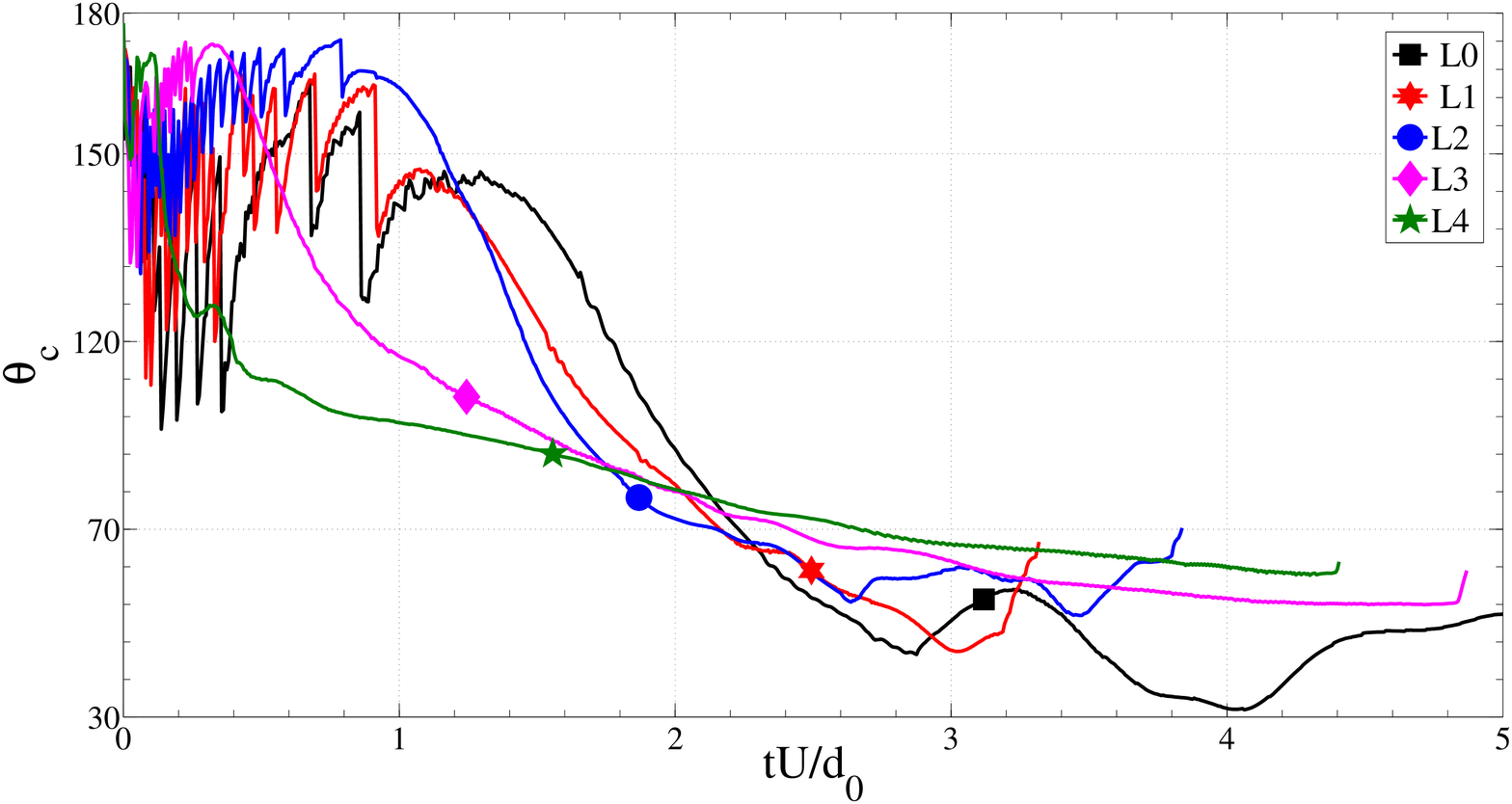}}}
\put(1.8,3.6){(a)}
\put(10.2,3.6){(b)}
\end{picture}
\end{center}
\caption{Computationally obtained dimensionless wetting diameter (a) and dynamic contact angle (b) with different points on the free surface using constant slip number $(\beta_\epsilon=30)$ for the cases in Table~\ref{tabconvergence2sph}. }
\label{ConvergencesphCS}
\end{figure*}

\begin{figure*}
\begin{center}
\unitlength1cm
\begin{picture}(12,3.5)
\put(-1.2,-2.35){\makebox(6,8){\includegraphics[width=0.5\textwidth]{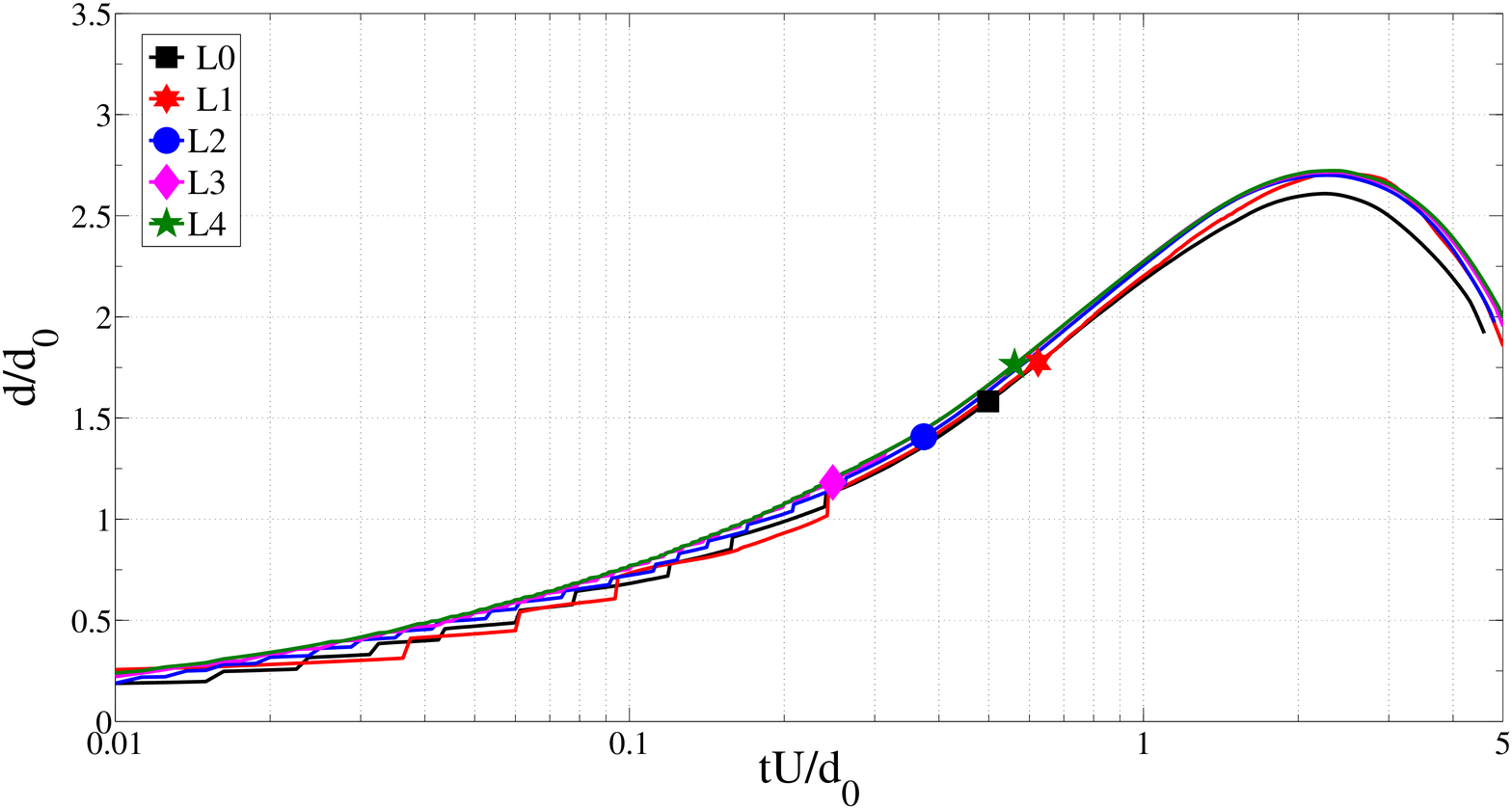}}}
\put(7.3,-2.35){\makebox(6,8){\includegraphics[width=0.5\textwidth]{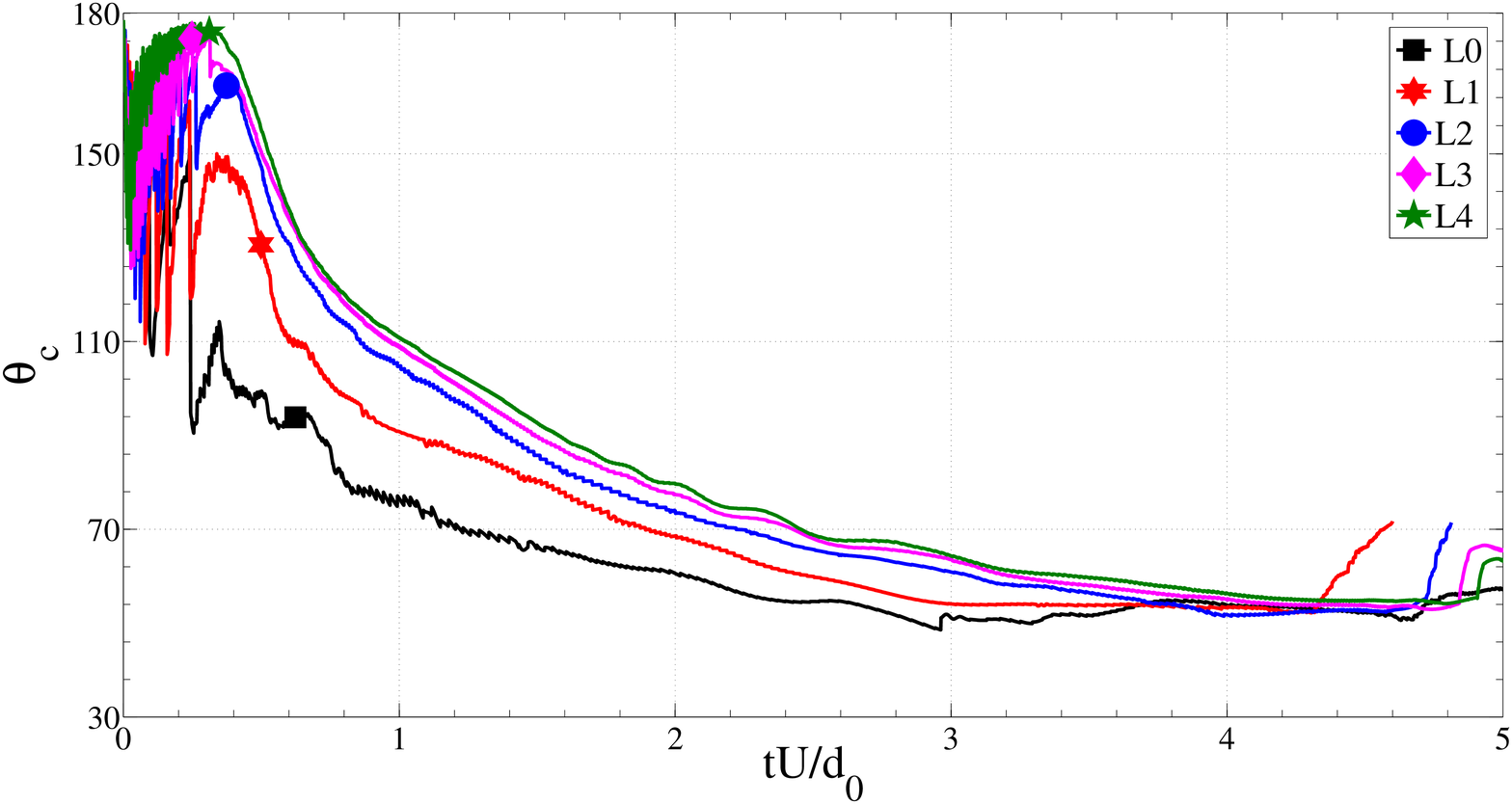}}}
\put(1.8,3.6){(a)}
\put(10.2,3.6){(b)}
\end{picture}
\end{center}
\caption{Computationally obtained dimensionless wetting diameter (a) and dynamic contact angle (b) with different points on the free surface using mesh dependent slip number $(\beta_\epsilon)$ for the cases in Table~\ref{tabconvergence2sph}. }
\label{ConvergencesphMDS}
\end{figure*}

We consider a spherical water droplet of diameter $d_0$~=~2.7~mm.
We take the characteristic length L~=~$d_0$/2~=~$r_0$, characteristic velocity U~=~$u_{imp}$  and the dimensionless numbers used in the computations are $\text{Re}$~=~1573, $\text{We}$~=~25, $\text{Fr}$~=~104 and $\theta_e$~=~75$^\circ$. 
Five variants for the free surface points have been used which are as follows: (i)~L0~:~25, (ii)~L1~:~50, (iii)~L2~:~100,  (iv)~L3~:~200 and (v)~L4~:~400.
First, we use a constant slip number $(\beta_\epsilon=30)$ in all the five variants.
From Figure~\ref{ConvergencesphCS}, we observe that the wetting diameter increases with increase in the number of points on the free surface.
Hence, we cannot obtain convergence using a constant slip number.
But from the wetting diameter curve, we can infer that the slip number has to be chosen in such a way that the wetting diameter is reduced with increase in the free surface points. 
Also, we know that the wetting diameter decreases with increase in the slip number value and the mesh size decreases with increase in the free surface points.
Hence, we need to use a mesh-dependent slip number.
Now, we perform computations using a mesh-dependent slip number, $\beta_\epsilon$~=~$\beta$/h$_0$, where h$_0$ is the initial size of the mesh on $\Gamma_F$.
For the values of slip number used in the computations, refer to Table~\ref{tabconvergence2sph}.
The computationally obtained wetting diameter and the dynamic contact angle are shown in Figure~\ref{ConvergencesphMDS}.
From Figure~\ref{ConvergencesphMDS}(a), we can observe that there is almost no influence of the free surface points on the wetting diameter. 
As $h_0$ tends to zero, $\beta_\epsilon$ tend to infinity which leads to the no-slip condition.
Hence, the slip number can be interpreted as an artificial friction/slip introduced in place of no slip condition for moving contact line problems. 
From Figure~\ref{ConvergencesphMDS}(b), we can observe that the free surface points have a significant influence on the dynamic contact angle. 
However, we can see convergence with L3 and L4 meshes.
Since, our aim is to accurately capture the flow dynamics of the droplet, in all the subsequent computations we use L3 mesh, i.e. 200 points on the free surface. 
\begin{table}
 \caption{Different cases of free surface points used for convergence study on a spherical liquid droplet} 
 \begin{center}
 \begin{tabular}{ccccl} 
 \hline
   Variant& Points on $\Gamma_F$ & h$_{0}$  & $\beta$ & $\beta_{\epsilon}$=$\frac{\beta}{h_{0}}$    \\
 \hline
 L0   & 25  & 0.12462872  & 0.467343  &  3.75 \\
 L1   & 50& 0.06231436 &  0.467343 & 7.5  \\
 L2  & 100 & 0.03115718    &  0.467343  & 15  \\
 L3   & 200 & 0.01557859    & 0.467343 & 30    \\
 L4   & 400 & 0.007789295    & 0.467343 & 60    \\
 \hline 
 \end{tabular}
 \end{center}
 \label{tabconvergence2sph}
 \end{table}
 
\subsection{Glycerin droplet}
In this section we consider glycerin droplets impinging perpendicularly on a smooth glass surface with equilibrium contact angle of $15^\circ$. 
The used values of physical parameters are : $\rho$~=~$1220$~kg~m$^{-3}$, $\mu$~=~$0.116$~N~s~m$^{-2}$ and $\sigma$~=~$0.063$~N~m$^{-1}$. 
Further, we take $\text{U}$~=~$u_{imp}$, $\text{L}$~=~$d_0$/2~=~$r_0$, $\beta_\epsilon$~=~$\beta$/h$_0$ with $h_0$~=~0.01557859  and $\text{g}$~=~$9.8$~m~s$^{-2}$.
The impinging velocity of the droplet is varied between $1.41$~m~s$^{-1}$ and $4.72$~m~s$^{-1}$. 
The obtained corresponding dimensionless numbers using the above parameters are given in Table~\ref{tabglycerin}. 
Computations are performed till the dimensionless time $100$ with a time step  length of $0.0005$.
For each case in Table~\ref{tabglycerin}, numerical simulations are performed with different slip numbers. 
The formation of secondary droplets (topological changes) is not considered and it is the reason for the choice of this specific range of impinging velocity of glycerin droplets.  

\begin{table}
\caption{Different cases of glycerin droplet used in this work} 
\begin{center}
\begin{tabular}{llllllc} 
\hline
   Case & $\text{Re}$  & $\text{We}$  & $\text{Fr}$ & $\text{u}_{imp}$(m~s$^{-1}$) & $\beta_\epsilon$ (identified) \\
\hline

A  &18& 47  &  166 &1.41 & 2000\\
B  & 24  &  81.5 & 286  & 1.854  & 750 \\
C  &31.5& 140 & 492 &2.43 & 300 \\
D   &37.5 & 201 & 706 & 2.91 & 200  \\
E  & 44.5  & 285.5 & 1002  & 3.47  & 125 \\
F   &61 & 528  &  1856  & 4.72  & 25\\
\hline 
\end{tabular}
\end{center}
\label{tabglycerin}
\end{table}

 \begin{figure*}
\begin{center}
\unitlength1cm
\begin{picture}(12,3.5)
\put(-1.2,-2.35){\makebox(6,8){\includegraphics[width=0.5\textwidth,height=0.25\textwidth]{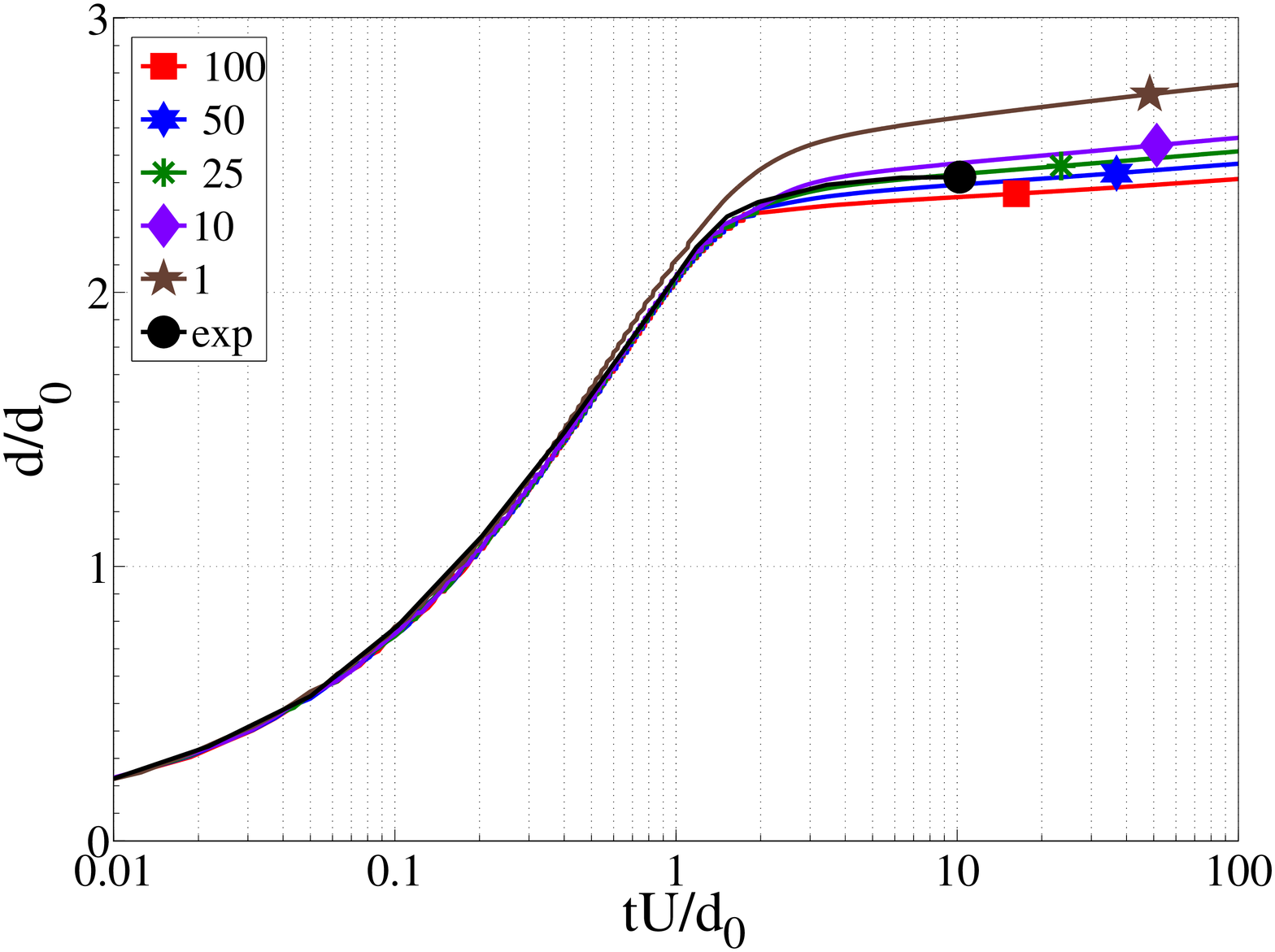}}}
\put(7.3,-2.35){\makebox(6,8){\includegraphics[width=0.5\textwidth,height=0.25\textwidth]{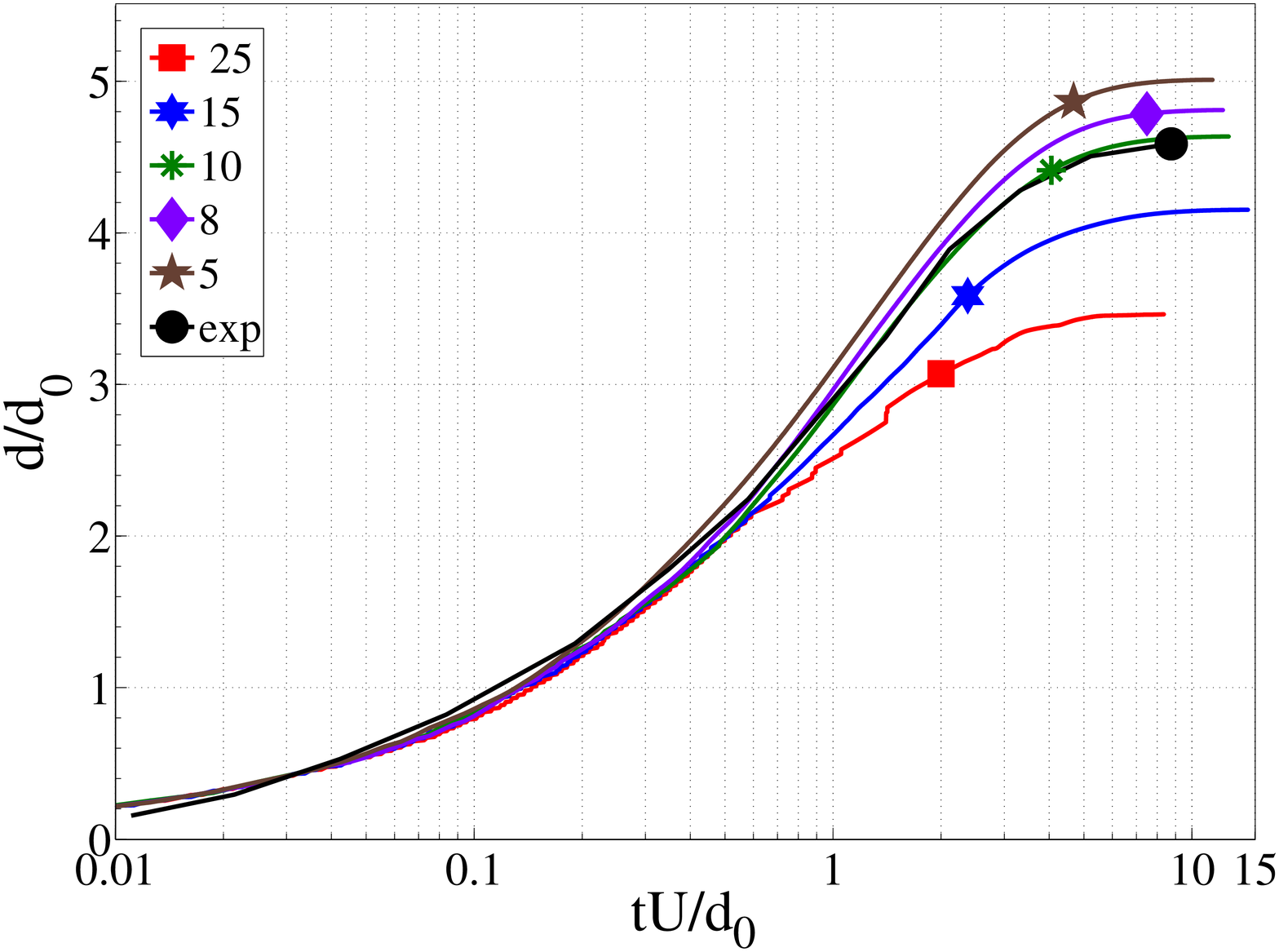}}}
\put(1.8,3.6){Case F}
\put(10.2,3.6){Case J}
\end{picture}
\end{center}
\caption{Computationally obtained dimensionless wetting diameter with different slip numbers for the glycerin droplet (Case F) and water droplet (Case J) is compared with experimental results. }
\label{expdrop}
\end{figure*}

We first study the influence of the slip number on the wetting diameter. 
Greater the value of slip number implies greater the effect of artificial friction. 
Hence, $\beta_\epsilon$ $\to$ $\infty$ implies no~slip and $\beta_\epsilon$ $\to 0$ implies free~slip. 
In Figure~\ref{expdrop}, the dimensionless wetting diameter obtained with different slip numbers for the case F is in good agreement with the experimentally observed values till the dimensionless time t~=~1, i.e., till the initial spreading phase of the droplet.
During the initial spreading phase, the effect of slip number on the flow dynamics is very minimal. 
However, after this initial phase different slip numbers induce different flow dynamics. 
For droplets with low slip numbers, the frictional resistance is less and hence the spreading velocity is higher when compared to the droplets with high slip numbers. 
Higher the spreading velocity, greater is the kinetic energy of the droplet. Also the wetting diameter directly depends on the kinetic energy of the droplet. 
Therefore, the maximum wetting diameter will be greater for low slip numbers and it can clearly be seen in Figure~\ref{expdrop}. 

The viscosity of glycerin is two orders higher than that of water. 
High viscosity of droplet induces a large resistant to the spreading and recoiling of droplet. 
Hence, the glycerin droplet deforms slowly on a smooth glass surface and it takes long time to attain its equilibrium wetting diameter. 
Generally, glycerin droplet does not rebound much due to high viscous dissipation.
Also the equilibrium contact angle will influence whether the droplet will recoil or not after reaching the maximum wetting diameter.   
The recoiling effect is not observed in the all the considered cases because the equilibrium contact angle is very small, i.e., $\theta_e=15^\circ$. 
The maximum wetting diameter is same as the final equilibrium wetting diameter in whole range of the investigated impinging velocities.
Also the maximum wetting diameter increases with increase in the impinging velocity of the droplet. 
We can observe in Figure~\ref{expdrop} that the slip numbers have a significant influence on the flow dynamics of droplet after the initial spreading phase. 
Hence, choosing an appropriate value for slip number in the computations is very essential indeed. 
On comparing the numerical simulations with experimental results from Sikalo~\cite{SIK03}, we identified an appropriate value for the slip number in each test case. 
The identified values of slip number ($\beta_{\epsilon}$) are 2000, 750, 300, 200, 125 and 25 for the cases A, B, C, D, E and F, respectively, and are  presented in Table~\ref{tabglycerin}. 
Note that all the slip number ($\beta_{\epsilon}$) values indicated above are of the form $\beta_\epsilon$~=~$\beta$/h$_0$ with $h_0$~=~0.01557859.
We can also observe that the identified values for the slip number decreases when the impact velocity increases for glycerin droplet.   

\subsection{Water droplet}
 
In this section we consider a water droplet impinging perpendicularly on a smooth glass surface with equilibrium contact angle of $10^\circ$. 
The used values of physical parameters are: $\rho$~=~$996$  kg~m$^{-3}$, $\mu$~=~$10^{-3}$~N~s~m$^{-2}$ and $\sigma$~=~$0.073$~N~m$^{-1}$. 
The impinging velocity of the water droplet is varied between $0.764$~m~s$^{-1}$ and $2.96$~m~s$^{-1}$.
The corresponding dimensionless numbers obtained using the above parameters are given in Table~\ref{tabwater}. 
Computations are performed till the dimensionless time $10$ with a time step length of $0.0005$.
For each case in Table~\ref{tabwater}, numerical simulations are performed with different slip numbers. 
Although the water droplet has comparable initial droplet diameter, equilibrium contact angle, surface tension and density to that of the glycerin droplet, its viscosity is two orders lower than that of glycerin.
Due to its low viscosity, the droplet spreads more than that of glycerin. 
The rate at which water spreads is much higher compared to glycerin and this is the reason we have performed the computations only till dimensionless time t~=~10.
In certain cases, the computations are stopped due to the formation of secondary droplets (topological changes) or due to dry out of the droplet  on $\Gamma_S$ at the axis of symmetry. 
Because of low viscosity of water, we have chosen an even lesser range of impinging velocity for water droplet in this study in order to resist the early formation of secondary drops or the occurrence of splashing.  

\begin{table}
 \caption{Different cases of water droplet used in this work} 
 \begin{center}
 \begin{tabular}{llllllc} 
 \hline
  Case & $\text{Re}$  & $\text{We}$  & $\text{Fr}$ & $\text{u}_{imp}$(m~s$^{-1}$) & $\beta_\epsilon$ (identified)  \\
 \hline
 
 G     &915& 9.5  &  50 &0.764 & 100\\
 H     & 1573  & 25 & 104  & 1.17  & 30\\
 I  &1820&38&196&1.52  & 20\\
 J   &2810 & 80.5 & 330 & 2.09 & 10\\
 K  & 2910  & 97 & 502  & 2.429  & 7\\
 L   &3545 & 144  &  746  & 2.96 & 4\\
 \hline
 \end{tabular}
 \end{center}
 \label{tabwater}
 \end{table}

\begin{figure*}
\begin{center}
\unitlength1cm
\begin{picture}(20,22.8)

\put(3.0,-1.2){\makebox(3,6){\includegraphics[trim=0cm 0cm 0cm 9cm, clip=true,width=8.5cm]{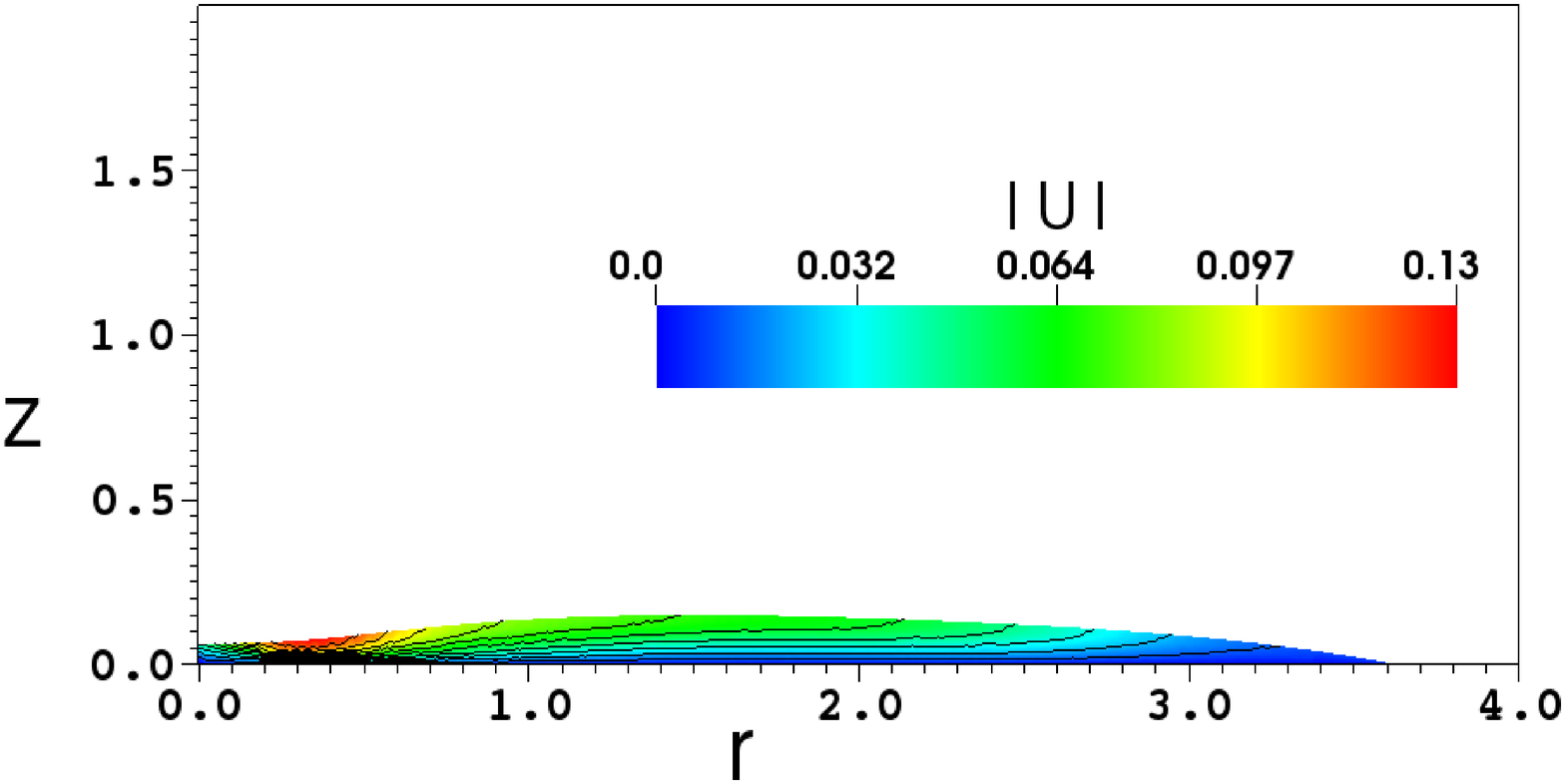}}}
\put(3.0,3.3){\makebox(3,6){\includegraphics[trim=0cm 0cm 0cm 9cm, clip=true,width=8.5cm]{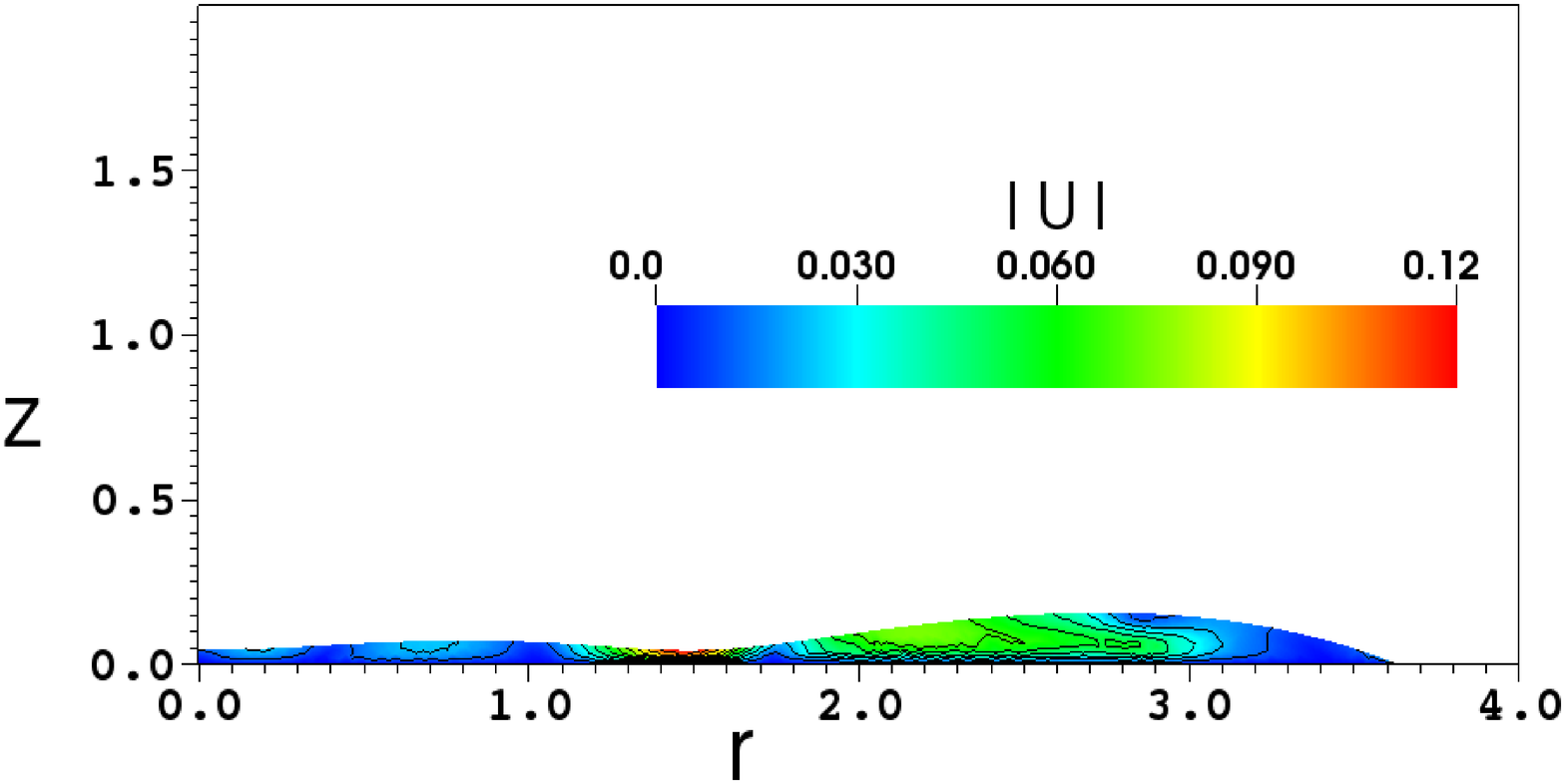}}}
\put(3.0,7.9){\makebox(3,6){\includegraphics[trim=0cm 0cm 0cm 9cm, clip=true,width=8.5cm]{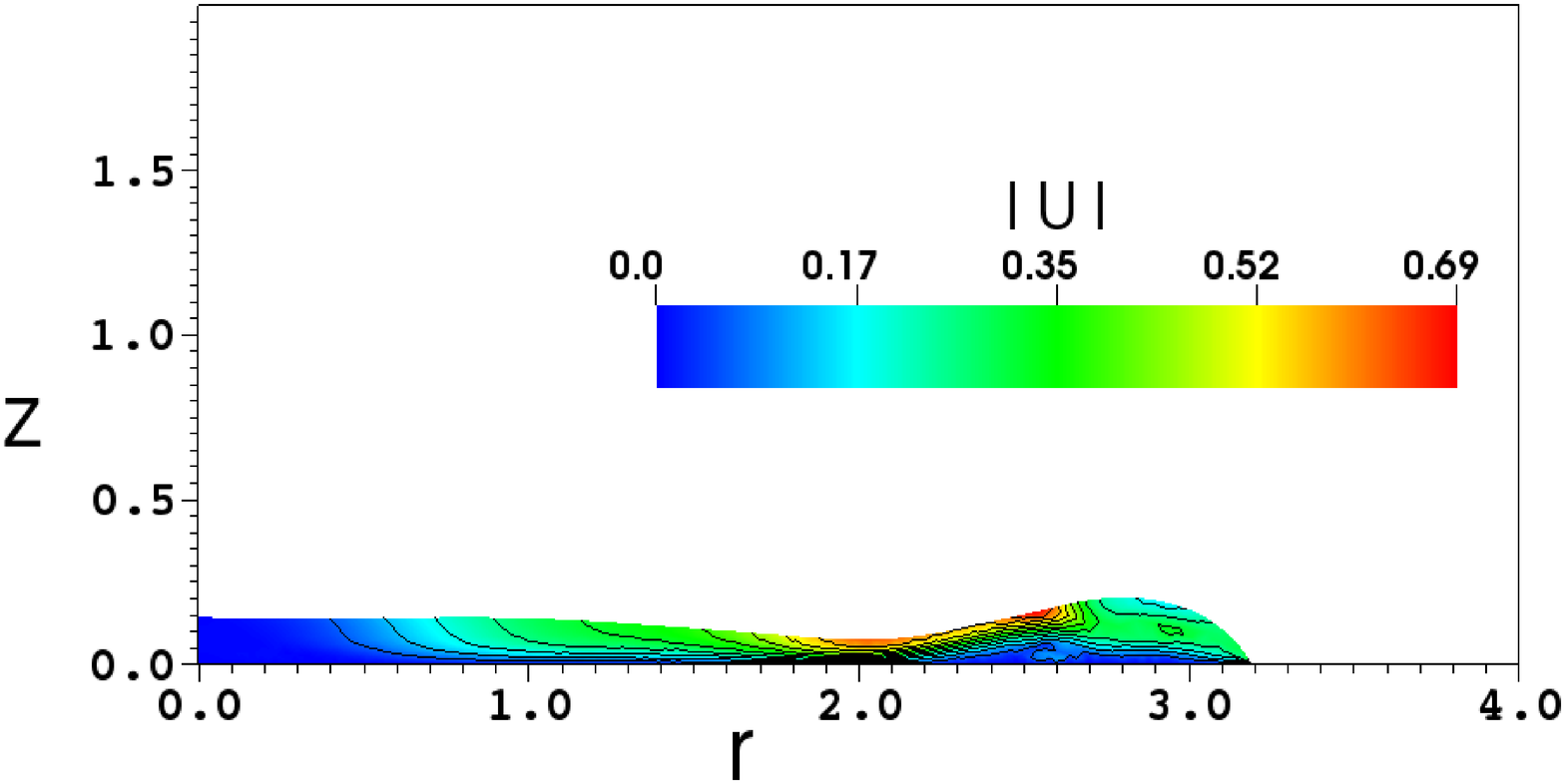}}}
\put(3.0,12.4){\makebox(3,6){\includegraphics[trim=0cm 0cm 0cm 9cm, clip=true,width=8.5cm]{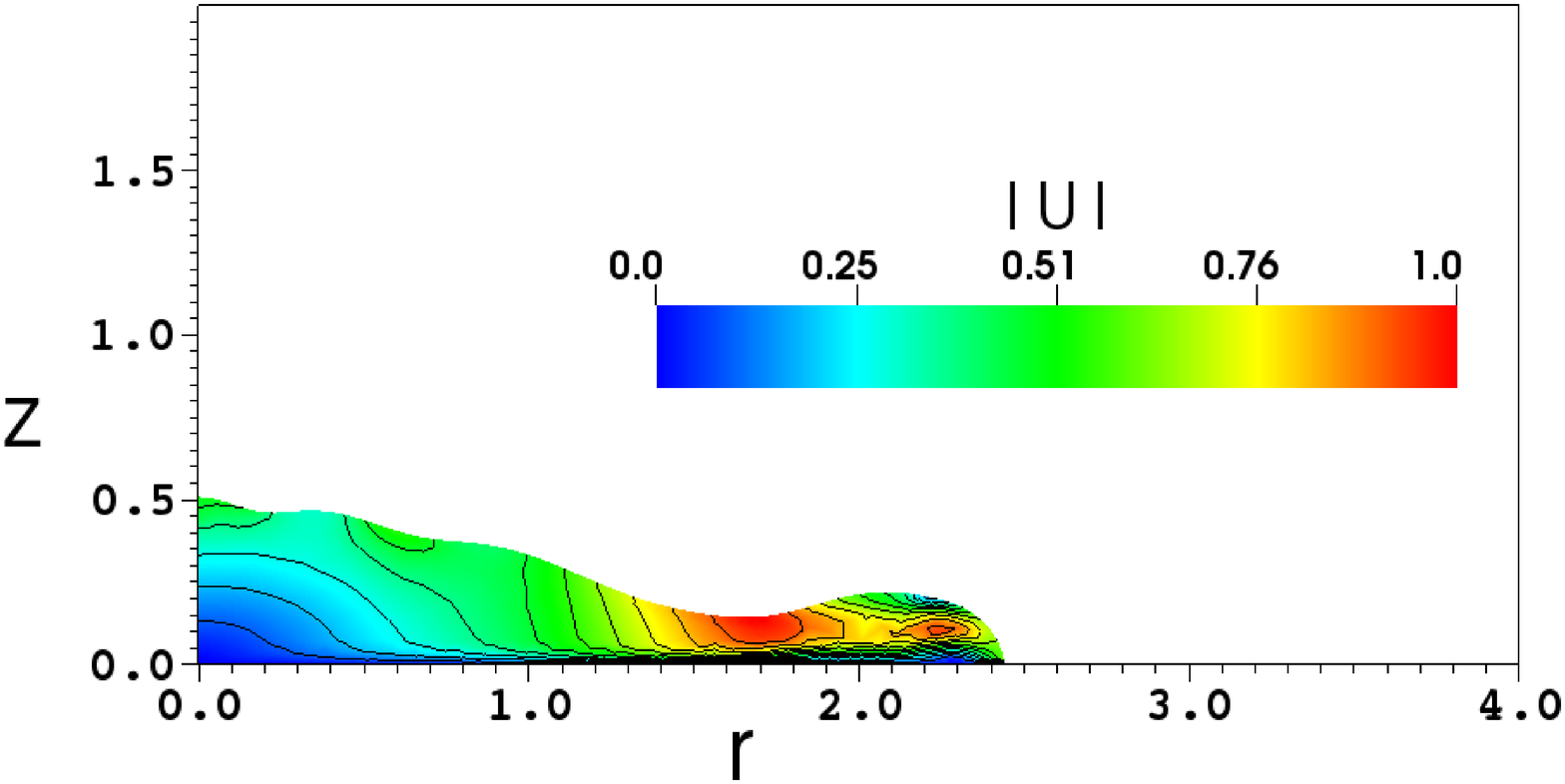}}}
\put(3.0,16.9){\makebox(3,6){\includegraphics[trim=0cm 0cm 0cm 9cm, clip=true,width=8.5cm]{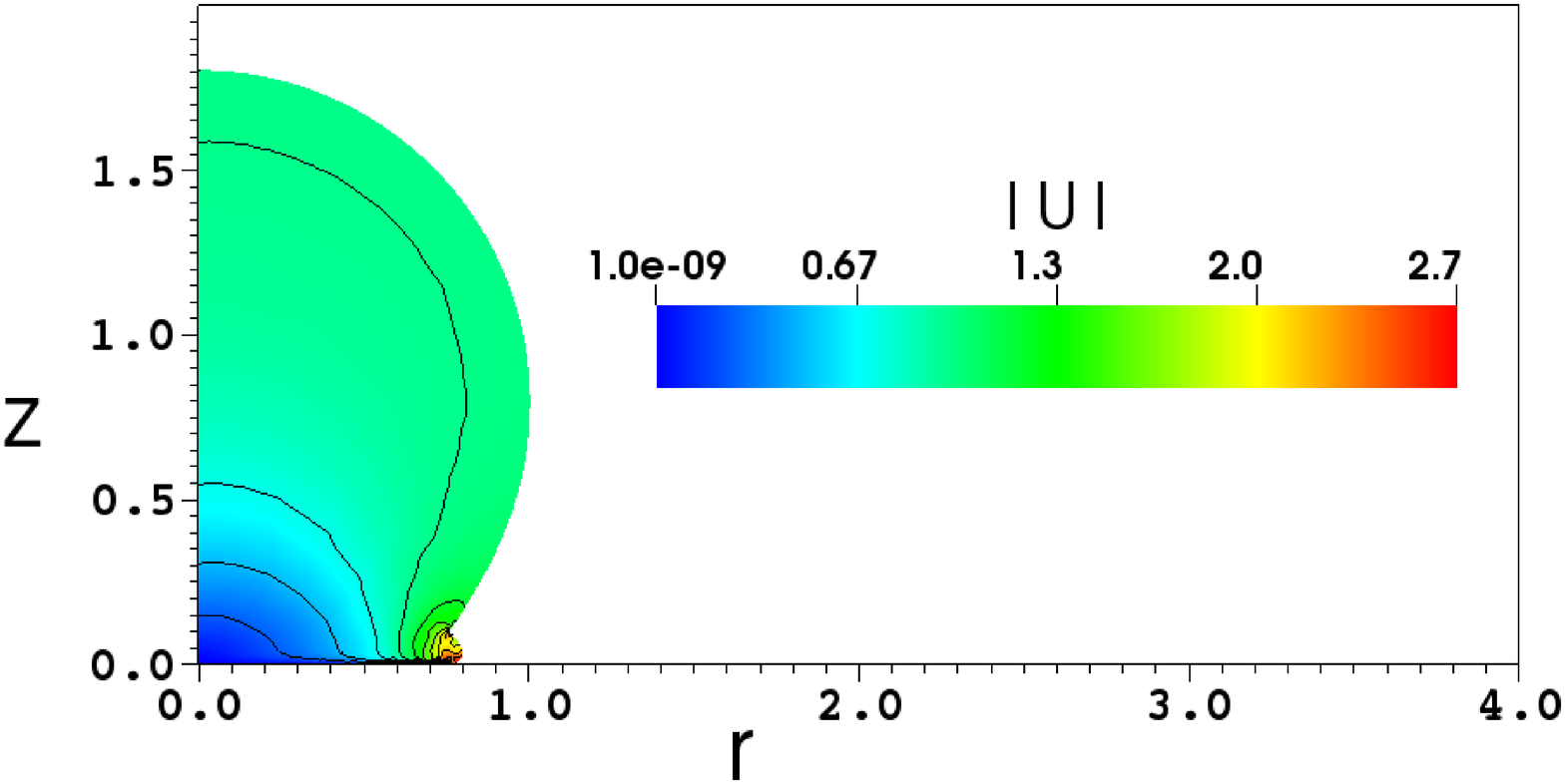}}}

\put(11.5,-1.2){\makebox(3,6){\includegraphics[trim=0cm 0cm 0cm 9cm, clip=true,width=8.5cm]{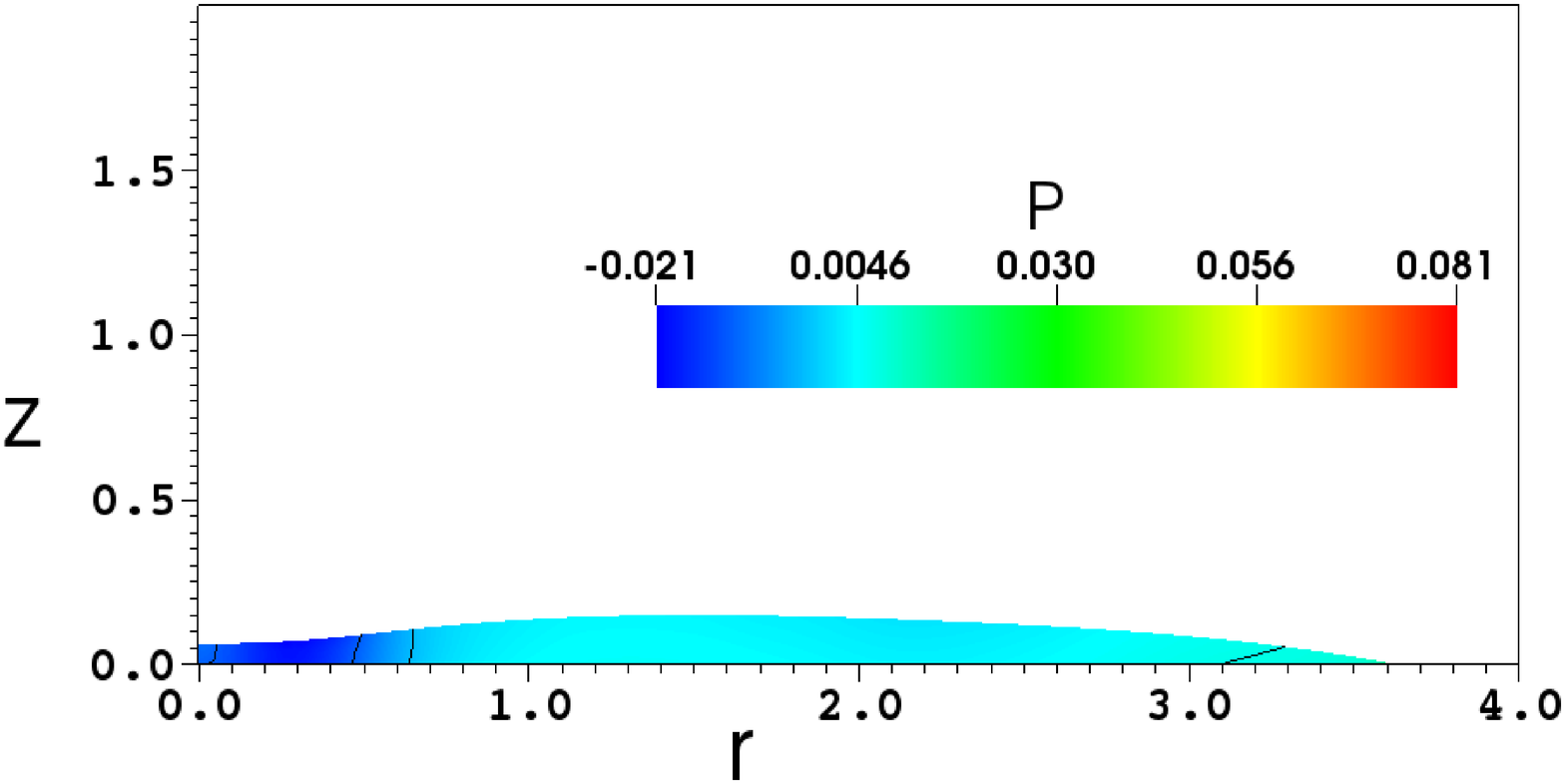}}}
\put(11.5,3.3){\makebox(3,6){\includegraphics[trim=0cm 0cm 0cm 9cm, clip=true,width=8.5cm]{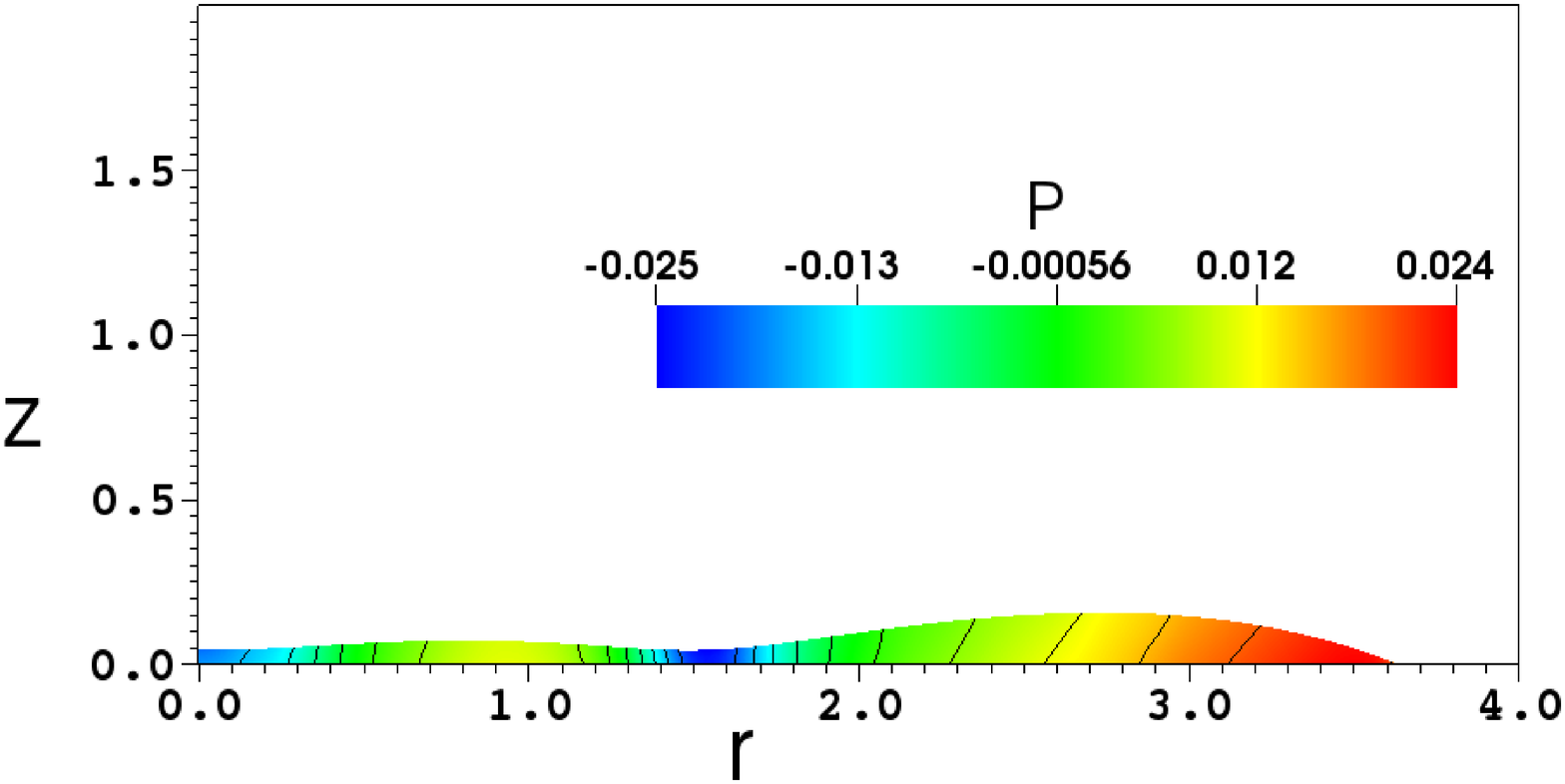}}}
\put(11.5,7.9){\makebox(3,6){\includegraphics[trim=0cm 0cm 0cm 9cm, clip=true,width=8.5cm]{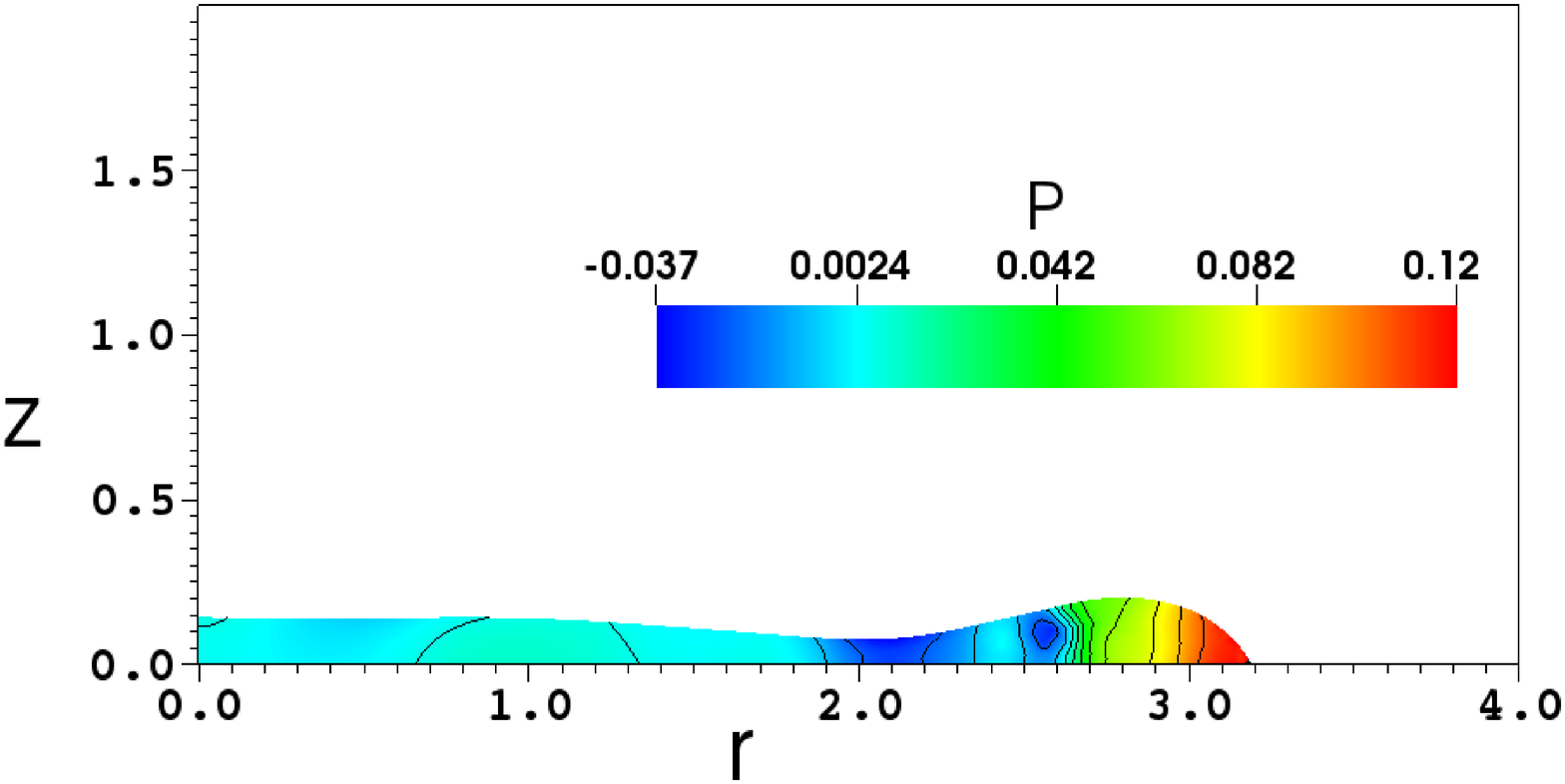}}}
\put(11.5,12.4){\makebox(3,6){\includegraphics[trim=0cm 0cm 0cm 9cm, clip=true,width=8.5cm]{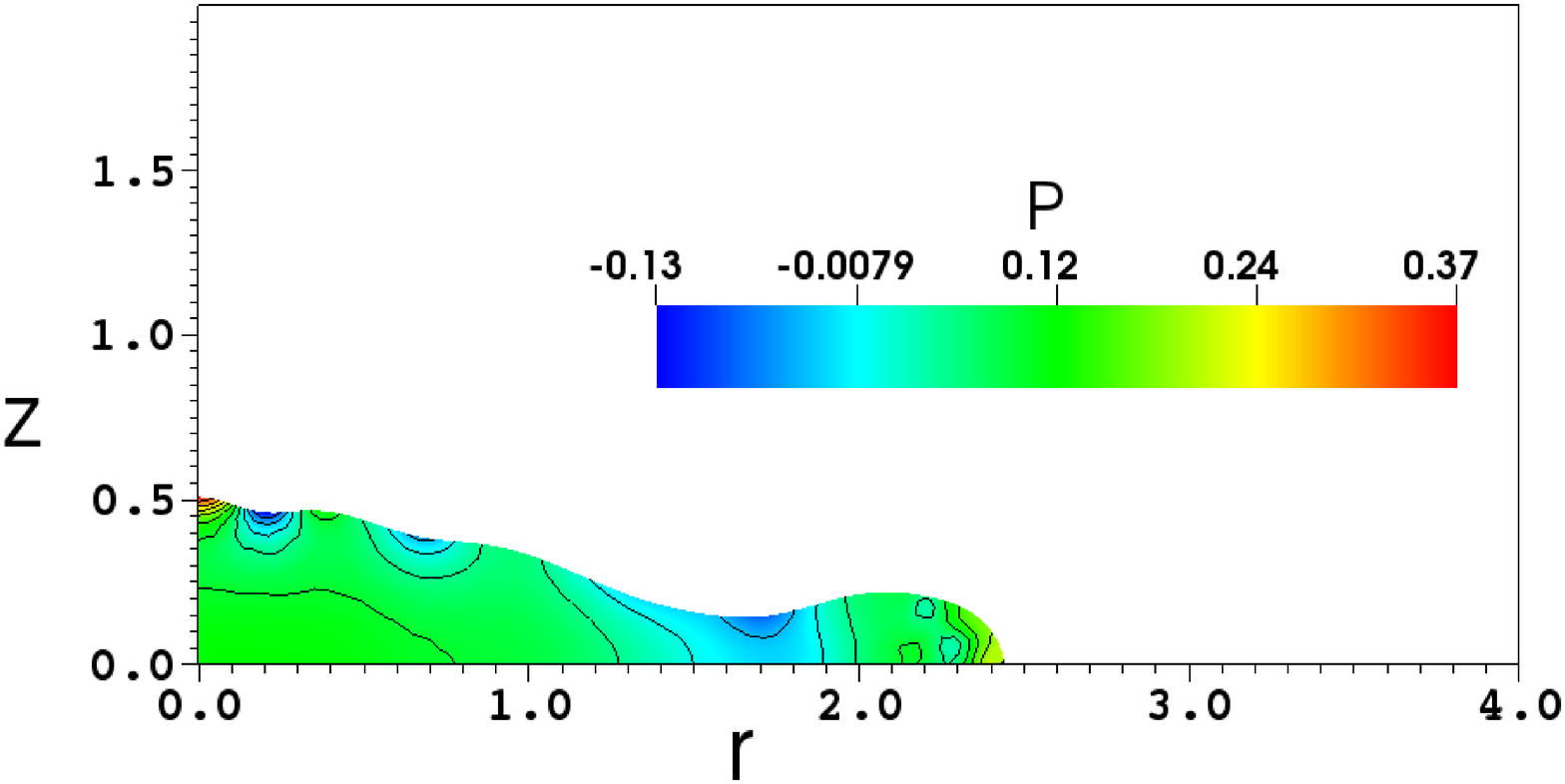}}}
\put(11.5,16.9){\makebox(3,6){\includegraphics[trim=0cm 0cm 0cm 9cm, clip=true,width=8.5cm]{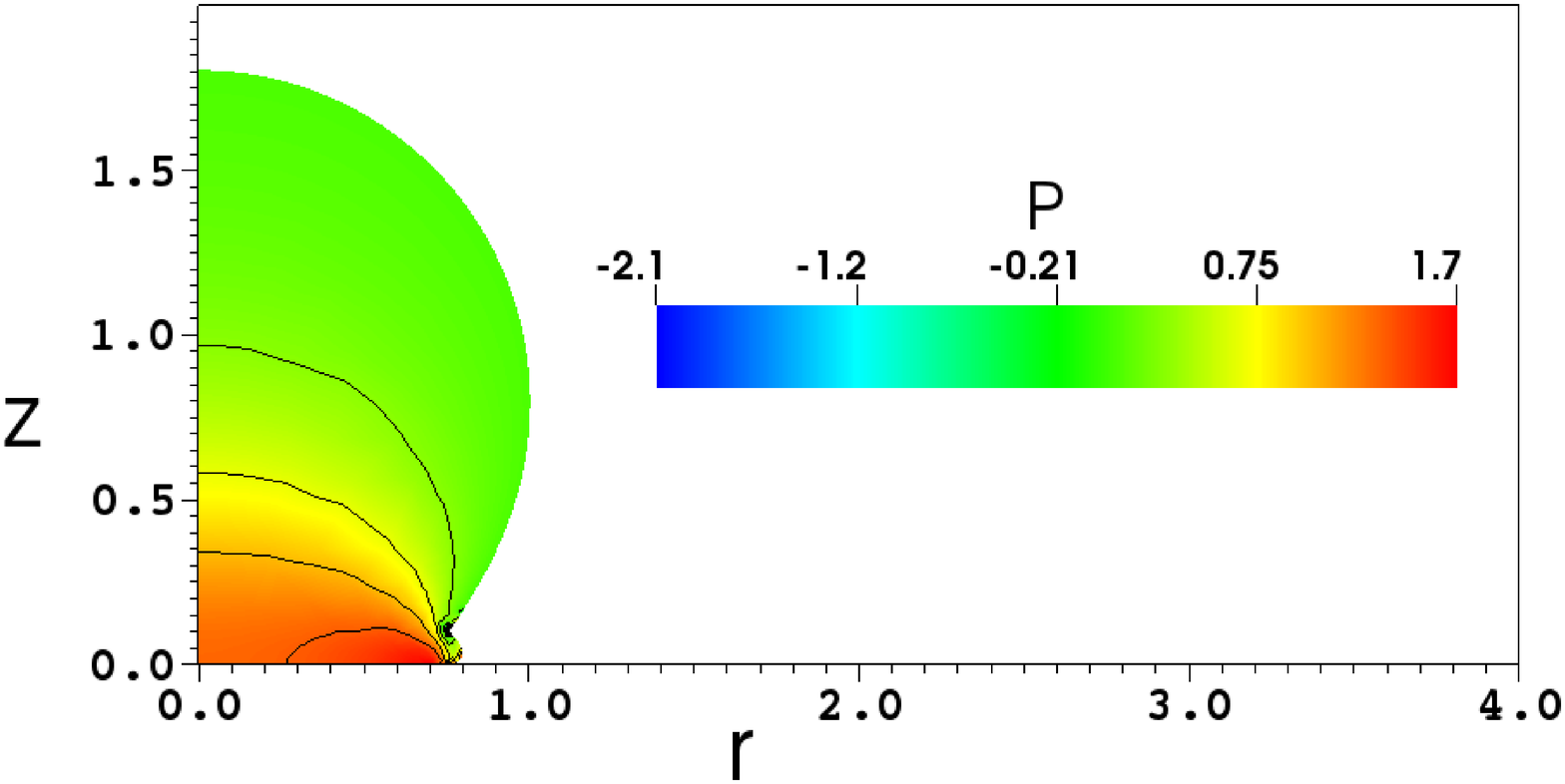}}}
\put(5.0,21.5){(a)}
\put(13.4,21.5){(b)}
\end{picture}
\end{center}
\caption{Magnitude of the velocity (a) and the pressure (b) contours of a impinging droplet (Case H in Table~\ref{tabwater}) at dimensionless times t~=~0.1, 1, 2, 5 and 10 from the top.}
\label{contour}
\end{figure*}

The numerical result for the case J in Table~\ref{tabwater} is shown in Figure~\ref{expdrop}.
During the initial spreading, we can observe a significant influence of the slip number on the flow dynamics. 
This is in total contrast to what we observed in the glycerin droplet.  
This can be attributed to the fact that water spreads swiftly compared to glycerin because of significantly lower viscosity. 
For a given impinging velocity, the wetting diameter is higher for low slip numbers which was also the case with glycerin droplet. 
Also with increase in impinging velocity, the wetting diameter of the spreading droplet increases. 
The recoiling effect is not observed because of the choice of a small equilibrium contact angle, i.e., $\theta_e$~=~$10^\circ$. 
From Figure~\ref{expdrop}, we observe that slip numbers have a significant influence on the flow dynamics of the water droplet. 
On comparing the numerical simulations with the experimental results from Sikalo~\cite{SIK03} and Roux et al.~\cite{ROUX04}, we identified an appropriate value for the slip number for each test case. 
The identified values of the slip number$(\beta_{\epsilon})$ are 100, 30, 20, 10, 7 and 4 for the cases G, H, I, J, K and L, respectively, and are presented in Table~\ref{tabwater}. 
We can also observe that the identified value for the slip number decreases when the impact velocity increases for water droplet which was also observed in glycerin droplet.   
On comparing the slip numbers for glycerin and water droplets with comparable impinging velocities, the slip numbers for glycerin droplets are almost two order higher than that of water droplet.
Figure~\ref{contour} depicts the magnitude of the velocity and the pressure contours of an impinging droplet (Case H in Table~\ref{tabwater}) at dimensionless time instances t~=~0.1, 1, 2, 5 and 10.

\subsection{Relation for the slip number}
Slip is a crucial factor in spreading of moving contact line problems.
The numerical method introduces a slip at the discrete level, effectively introducing slip length on the order of the mesh size.
Several authors~\cite{MORIA, REN01, WE08, ZAL09} have reported a convergence breakdown with the grid refinement and they overcame this by using a mesh-dependent slip for numerical solutions of moving contact line problems, which we observed in the earlier mesh convergence study.
A relation between the Greenspan slip coefficient and the grid-spacing of the numerical scheme has been proposed by Moriarty et al.~\cite{MORIA} using curve fitting for the moving contact line problem arising in dry wall coating.
Hence, this gives us the motivation to find a relation for the slip number applicable to impinging droplets.
In the previous sections, we identified appropriate slip values for several test cases of glycerin and water droplet impinging on a glass surface. 
The dynamics of wetting for glycerin and water are not the same, e.g. different time scales for reaching maximum wetting diameter which is due to different viscosity in both liquids.
However, the whole area of dynamic wetting has been motivated by developing models which are capable of describing widely varying wetting phenomena with the same set of parameters.
Hence, this motivates us to obtain a relation for the slip number applicable to any liquid. 

We have studied the influence of the slip number on the flow dynamics using the dimensionless wetting diameter which is also known as spread factor.
The spreading behavior largely depends on the viscous and capillary forces of the droplet. 
The dimensionless numbers which account for these forces are the Reynolds and the Weber number, respectively. 
From the slip values, we observe that with increase in the Reynolds number, the slip number decreases and the decrease is quite rapid indicating that the relation may not be linear but could be exponential. 
The same behavior is also observed with the Weber number. 
Both the Reynolds and the Weber number play a major role in determining the spreading behavior. 
The dimensionless number which represents the relative effect of the viscous forces and surface tension is the capillary number, i.e., the ratio of Weber number to Reynolds number.
However, trying to find a relation between capillary number and slip number will lead to the assumption that the relative effect of viscous forces and surface tension would be the same for all the droplets, which may not be true always.
Hence, using the identified slip values for several test cases, we obtain a relation for slip number in terms of the mesh size, the Reynolds and the Weber number. 
For curve fitting, we used an online package called ``Labfit''. 
Upon fitting, we have obtained the following relation.
\begin{equation}\label{sliprelation}
        \beta_{\epsilon} =    \frac{\beta}{h_{0}}, \qquad      \beta  = \alpha {\text{Re}}^{\gamma} + \lambda {\text{We}}^{\delta},
\end{equation}
where 
\begin{align*}
 &\alpha = 4.796842276577\times10^{5}, \qquad
& \gamma = -3.339370111853,\\
 &\lambda = 2.021796892969\times10^{1}, \qquad
&\delta = -1.142224345078.
\end{align*}
Note that we have used L~=~$r_0$ in computations and the fit is using the Reynolds and Weber number which also are in terms of L~=~$r_0$.
However, in the literature authors have used L~=~$d_0$. In such cases, the slip number shall be used as : $\beta_\epsilon$=$\beta$/2h$_0$, where $\beta$ is obtained from the proposed relation~(\ref{sliprelation}). 

\subsection{Validation of the proposed slip relation}
\begin{figure*}
\begin{center}
\unitlength1cm
\begin{picture}(12,3.5)
\put(-1.2,-2.35){\makebox(6,8){\includegraphics[width=0.5\textwidth]{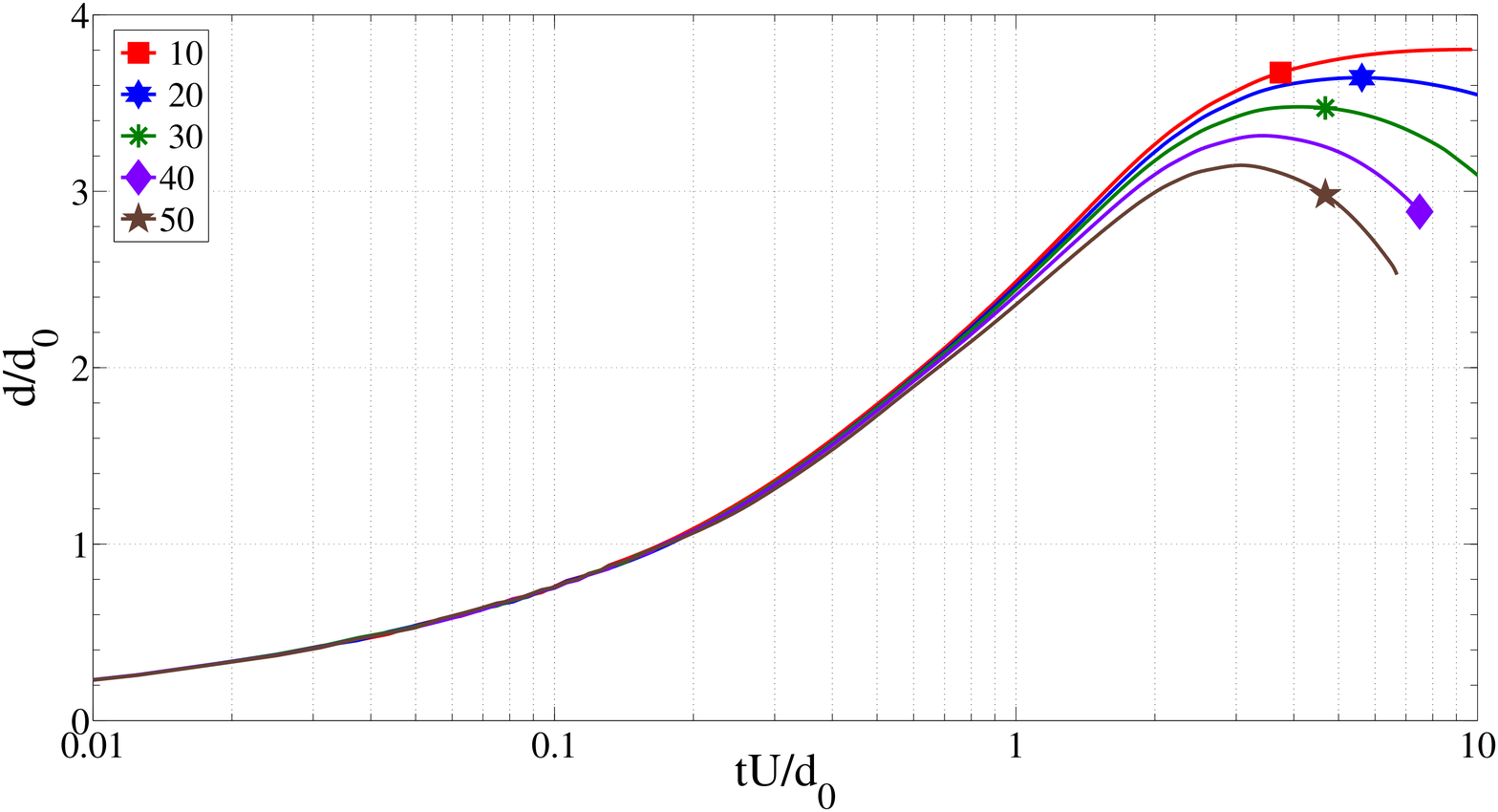}}}
\put(7.3,-2.35){\makebox(6,8){\includegraphics[width=0.5\textwidth]{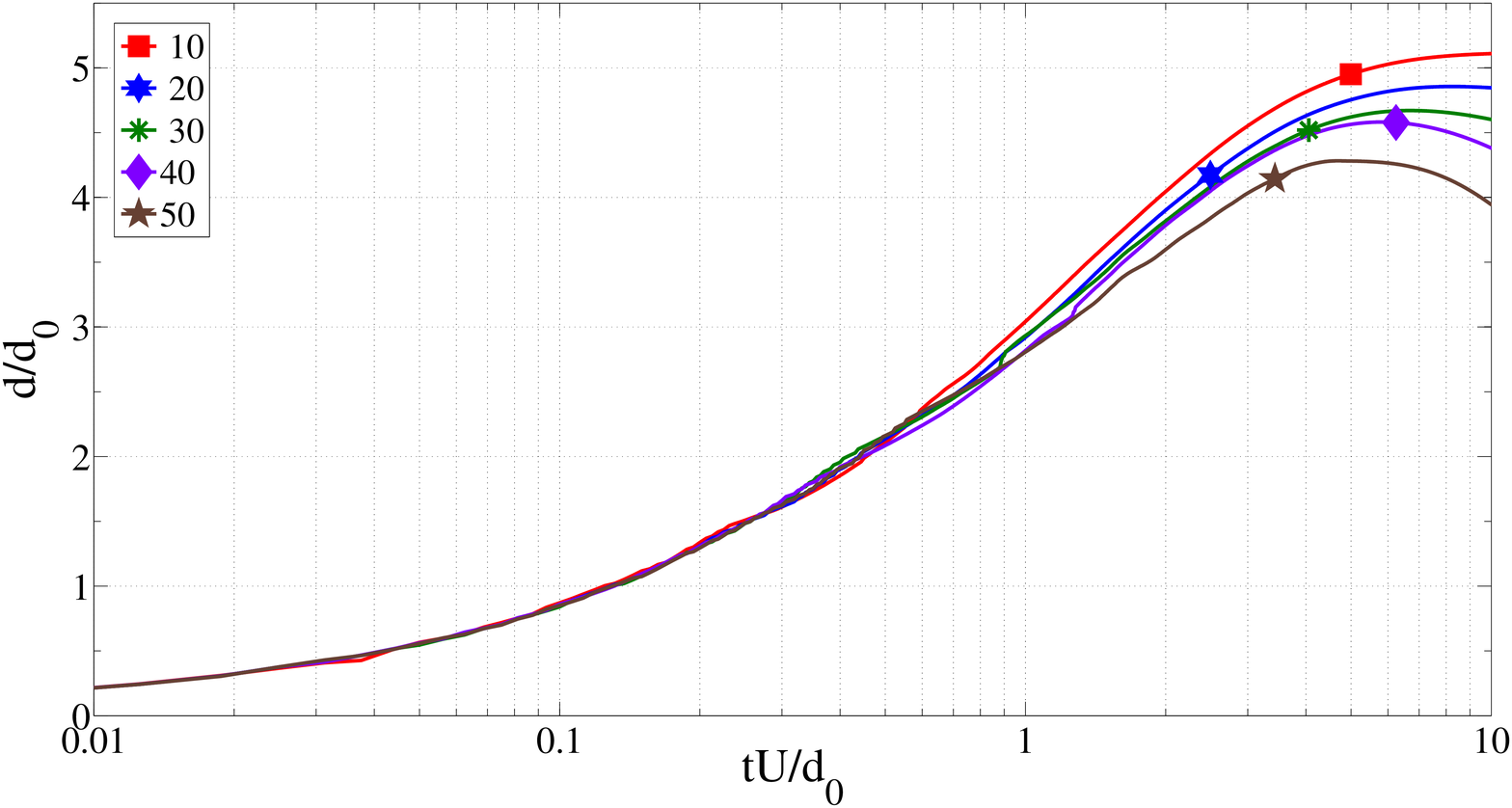}}}
\put(1.8,3.6){(a)}
\put(10.2,3.6){(b)}
\end{picture}
\end{center}
\caption{Computationally obtained dimensionless wetting diameter with different equilibrium contact angles for the cases H and L in Table~\ref{tabwater}. }
\label{WaterCApics1}
\end{figure*}

In this section we perform an array of computations by varying the Reynolds number, Weber number and the equilibrium contact angle to validate the proposed relation for the slip number.  
To validate the relation for any hydrophilic surface, we study the influence of contact angle on the flow dynamics of impinging droplet. 
We consider the cases H and L in Table~\ref{tabwater}. 
We perform computations for these two cases with the respective slip values as predicted by the proposed relation~(\ref{sliprelation}) and by varying the equilibrium contact using five variants: (i) 10$^\circ$, (ii) 20$^\circ$, (iii) 30$^\circ$, (iv) 40$^\circ$ and (v) 50$^\circ$. 
From Figure~\ref{WaterCApics1}, we can observe that the effect of contact angle on the flow dynamics is quite significant for both the flows. 
However, we can predict the maximum dimensionless wetting diameter for flows with varying contact angles using the following analytical relation, refer~\cite{UKI05}.
\begin{align}\label{analyticalexp}
(\text{We}+12) {\text{Wd}}_{\text{A}} = 8 + {\text{Wd}}_{\text{A}}^3 \left[3(1-\text{cos}\theta) + 4\frac{\text{We}}{\sqrt{\text{Re}}} \right]
\end{align}

\begin{table}
 \caption{Different cases of equilibrium contact angles for Water droplet with Re~=~1573} 
 \begin{center}
 \begin{tabular}{cccccc} 

 \hline
    $\theta_e$ & ${\text{Wd}}_{\text{N}}$& ${\text{Wd}}_{\text{A}}$ & Relative error ($\%$)   \\
 \hline
 
 10  &3.8037& 4.0774  &  6.71  \\
 20 & 3.6435  &4.0018 & 8.95   \\
 30 &3.4783 & 3.8868  &10.51  \\
 40  &3.3148 & 3.7453 & 11.49  \\
 50  &3.1469  & 3.5901 & 12.34   \\
 70 & 2.7945 & 3.2789 & 14.77 \\
 90 & 2.5280 & 3.0062 & 15.91  \\
 100 & 2.4476 & 2.8911 & 15.34 \\
 120 & 2.3031 & 2.7055 & 14.87 \\
 140 & 2.1905 & 2.5777 &  15.02 \\
 \hline 

 \end{tabular}
 \end{center}
 \label{maxwdcompare3146}
 \end{table}

\begin{table}
 \caption{Comparison of numerical and experimental results for validation} 
 \begin{center}
 \begin{tabular}{llllllllll} 

 \hline
    \re & \we & \fr & $\theta_e$ & $\beta_\epsilon$  &  $\text{Wd}_{\text{E}}$ & $\text{Wd}_{\text{N}}$ & Er   \\
 \hline
 
 1042  &29.5& 2257  &  27 & 27.19 & 3.47 & 3.45  &0.58 \\
 1649 & 59  & 2846 & 27 & 12.32 & 4.07 & 4.07 & 0 \\
 2129 & 85.5 & 3163  &27 & 8.06 & 4.2 & 4.39 & 4.52 \\
 2528.5  & 109.5 & 3342 & 27 & 6.08 & 4.3 & 4.6 &6.98 \\
 1042  & 29.5  & 2257 & 62  & 27.19 & 3.15 & 2.91 & 7.62 \\
 1649 & 59 & 2846 & 62  & 12.32 & 3.56 & 3.54 & 0.56 \\
 2129 & 85.5 & 3163 & 62  & 8.06 & 3.82 & 3.89 & 1.83 \\
 2528.5 & 109.5 & 3342 & 62 & 6.08 & 4.1 & 4.1 & 0 \\

 \hline 

 \end{tabular}
 \end{center}
 \label{maxwdcompareexp}
 \end{table}

The maximum wetting diameter obtained numerically $(\text{Wd}_{\text{N}})$ from these simulations are compared with the values predicted by the analytical expression$(\text{Wd}_{\text{A}})$ in Table~\ref{maxwdcompare3146}.
We have performed the simulations for wetting and partially wetting liquids.
It has also been established that the mean error in predicting the maximum wetting diameter by the using the analytical expression is 5.09$\%$  with a standard deviation of 5.05$\%$.
For the case with equilibrium contact angle of $10^\circ$, we have a relative error of 6.71$\%$. 
However, this is the case we had obtained the slip number based on comparison with experiments.
We assume that the experimental results are accurate and hence we have a error in the maximum wetting diameter predicted by analytical expression to be 6.71$\%$.
In this case, the analytical expression over-predicts when compared to experimental results.
Even though the relative error in most of cases in Table~\ref{maxwdcompare3146} is more than 10$\%$, due to over-prediction of the analytical expression we expect the relative error to be less than 10$\%$ for the cases with equilibrium contact angles $\theta_e < 90^\circ$, as our calibration of slip number is based on the experiments.
For hydrophobic and super-hydrophobic surfaces, i.e. for $\theta_e > 90^\circ$, the proposed relation may not be valid which could be a future scope for research. 
Hence, we can use the obtained correlation for the slip number values for droplet impinging on a hydrophilic surface. 

We have used experimental data from~Sikalo~\cite{SIK03} and Roux et al.~\cite{ROUX04} to compare the numerical results and derive the relation for slip number.
We now compare the numerical results obtained using the proposed slip relation~(\ref{sliprelation}) with some other experimental data provided in Ford et al.~\cite{FORD67}
The considered test cases are indicated in Table~\ref{maxwdcompareexp}. 
Note that we have used $h_0$~=~0.01557859 in the computations and we have considered only droplet impinging on a hydrophilic surface.
From the Table~\ref{maxwdcompareexp}, we can observe that the relative error (Er) in the maximum wetting diameter between the experimental and the numerical result is safely less than $10\%$ for all cases. 
This further validates the proposed relation for the slip number for hydrophilic surfaces.

\section{Summary and Future Work}
\label{s5}
In this paper we proposed a free surface mesh-dependent relation~(\ref{sliprelation}) for the slip number used in the Navier-slip with friction boundary condition on the liquid-solid interface for computations of liquid droplet impinging on a hydrophilic surface.
An array of numerical simulations of liquid droplet impinging on a horizontal surface are presented in the paper.
Finite element simulations are performed using arbitrary Lagrangian-Eulerian approach to study the effect of slip number on the flow dynamics of glycerin  and water droplet impingement.  
Computations are performed  for different impact velocities and droplet sizes.
Appropriate value for the slip number in each test case is identified by comparing the numerical results with experiments. 
Further, using the identified slip numbers for the given Reynolds, Weber number and the  mesh size, a relation is derived for the slip number.
The proposed relation is then validated by comparing the computationally obtained maximum wetting diameter with the analytical predictions and other experiments.
The proposed relation is more reliable for droplet impinging on a hydrophilic surface.
Moreover, for droplet impinging on hydrophobic and super-hydrophobic surfaces, the same relation for slip number may not be appropriate.
However, this could still give a good indication of the range of the slip number to be used in computations.
Further research has to be done for the choice of exact slip number for droplet impinging on hydrophobic and super-hydrophobic surfaces.

\bibliographystyle{IEEEtran}
\bibliography{masterlit}

\end{document}